\documentclass[aip,reprint]{revtex4-1}

\usepackage{graphicx}
\usepackage{blindtext}
\usepackage{lipsum}
\usepackage{color}
\usepackage[dvipsnames]{xcolor}
\usepackage{amsmath}
\usepackage{amssymb}
\usepackage{epstopdf}
\usepackage{bm}
\usepackage{appendix}
\usepackage{pgffor}
\usepackage{physics} 
\usepackage{nicefrac}

\usepackage[utf8]{inputenc}
\usepackage[T1]{fontenc}
\usepackage{mathptmx}

\begin{document}
\preprint{}


\title{Dynamics of Uniaxial-to-Biaxial Nematics Switching in Suspensions of Hard Cuboids} 



\author{Effran Mirzad Rafael}
\affiliation{Department of Chemical Engineering and Analytical Science, The University of Manchester, Manchester, M13 9PL, United Kingdom}%
\author{Luca Tonti}
\affiliation{Department of Chemical Engineering and Analytical Science, The University of Manchester, Manchester, M13 9PL, United Kingdom}%
\author{Daniel Corbett}
\affiliation{Department of Chemical Engineering and Analytical Science, The University of Manchester, Manchester, M13 9PL, United Kingdom}%
\author{Alejandro Cuetos}
\affiliation{Department of Physical, Chemical and Natural Systems, Pablo de Olavide University, 41013 Sevilla, Spain}%
\author{Alessandro Patti}
\email{alessandro.patti@manchester.ac.uk}
\affiliation{Department of Chemical Engineering and Analytical Science, The University of Manchester, Manchester, M13 9PL, United Kingdom}%



\begin{abstract}
Field-induced reorientation of colloidal particles is especially relevant to manipulate the optical properties of a nanomaterial for target applications. We have recently shown that surprisingly feeble external stimuli are able to transform uniaxial nematic liquid crystals (LCs) of cuboidal particles into biaxial nematic LCs. In the light of these results, here we apply an external field that forces the reorientation of colloidal cuboids in nematic LCs and sparks a uniaxial-to-biaxial texture switching. By Dynamic Monte Carlo simulation, we investigate the unsteady-state reorientation dynamics at the particle scale when the field is applied (uniaxial-to-biaxial switching) and then removed (biaxial-to-uniaxial switching). We detect a strong correlation between the response time, being the time taken for the system to reorient, and particle anisotropy, which spans from rod-like to plate-like geometries. Interestingly, self-dual shaped cuboids, theoretically considered as the most suitable to promote phase biaxiality for being exactly in between prolate and oblate particles, exhibit surprisingly slow response times, especially if compared to prolate cuboids. 
\end{abstract}

\pacs{}

\maketitle 

\section{Introduction}
Colloids are biphasic systems comprising particles homogeneously dispersed in a medium. In colloidal suspensions, the dispersed phase consists of solid particles, while the continuous phase is a liquid. In particular, the dispersed particles should have at least in one direction a dimension roughly between 1 nm and 1 $\mu$m, so that gravitational and thermal forces compensate each other \cite{mormann2017preferred}. This balance allows the dispersed particles to remain suspended and to diffuse randomly \textit{via} Brownian motion, named after the Scottish botanist Robert Brown who, in 1827, described the persistent and casual jumpy moves of organelles suspended in water \cite{brown_2015}. If the suspended particles are anisotropic, under certain conditions, they can self-organise into liquid-crystalline phases. Liquid crystals (LCs) are mesophases that flow like liquids but, exhibit a significant degree of internal ordering like crystals. A common LC morphology is the uniaxial nematic ($\rm N_{U}$) phase where particles have one axis pointing collectively in the same direction, but their centres of mass are randomly distributed. This merely orientational ordering allows nematic LCs to exhibit optical birefringence while maintaining mechanical fluidity. Currently, LCs are deployed in a multitude of optical technology, including (but not limited to) commercial displays \cite{chen2018liquid}, displays for virtual augmented reality \cite{huang2018liquid} and even smart windows with memory displays \cite{dabrowski2018fluorinated}. The roll-out of these high-tech products are coupled alongside further advances to understand and control the morphology of LCs \cite{meyer2020biaxiality} as well as techniques to optimise how they are manufactured \cite{cousins2019squeezing}, highlighting the relevance of LCs in today's research landscape. Very recently, there has been reignited interest in the biaxial nematic phase ($\rm N_{B}$) and its potential to be incorporated into display technology. In contrast to the $\rm N_{U}$ phase, the $\rm N_{B}$ phase possesses two optical axes due to the alignment of the three  directors, making it very appealing for the design of nanomaterials with novel optical properties  \cite{luckhurst2015biaxial}. Equally important, the $\rm N_{B}$ phase is also foreseen to realise faster switching through its minor axis switching mode, an aspect that could potentially improve refresh rates in displays \cite{lee2007dynamics, berardi2008, ricci2015field}. Despite these promising features, the existence of stable molecular $\rm N_{B}$ phases is still an ongoing debate within the LC community. While a biaxial geometry is indeed necessary to observe $\rm N_{B}$ phases, it has been shown that this requirement might not be sufficient as the $\rm N_B$ phase tends to be metastable with respect to other phases, including $\rm N_U$ and smectic (Sm) LCs \cite{taylor1991nematic}. This is for instance the case of colloidal cuboids, which can only form $\rm N_B$ phases at sufficiently large size dispersity \cite{belli2011polydispersity, rafael2020self}, extreme anisotropy \cite{dussi2018hard}, in the presence of depletants \cite{belli2012depletion} or upon application of an external field \cite{op2014tuning, cuetos2019biaxial}. Unless at least one of these conditions are met, systems of monodisperse or bidisperse cuboids cannot form $\rm N_B$ phases \cite{cuetos2017phase, patti2018monte}. For the monodisperse case, simulation results showed excellent qualitative agreement with sedimentation experiments of highly uniform colloidal cuboids synthesised by Yang and co-workers, which also preclude the existence of stable $\rm N_{B}$ phases at equilibrium \cite{yang2018synthesis}. Research efforts have also been made to ascertain, by theory, simulation and experiments, the existence of $\rm N_B$ phases in systems of other biaxial particles \cite{chiappini2019biaxial, tasios2017simulation, van2009experimental, peroukidis2013supramolecular, peroukidis2013phase, peroukidis2014biaxial, skutnik2020biaxial, van2010isotropic, van2010onsager, peroukidis2020field} or mixtures of uniaxial particles \cite{cuetos2008thermotropic}.

Despite such a widespread interest in mapping the phase behaviour of colloidal suspensions of biaxial particles, including cuboids, the study of their dynamics is still at an embryonic stage, especially for the difficulty of finding suitable interaction potentials that could describe exotic geometries and still be framed within a simulation technique. Exotic shapes are commonly described by hard-core potentials, but these cannot be directly employed in Brownian dynamics (BD) or Molecular Dynamics (MD) simulations. Nevertheless, it is only by assessing the dynamics that one will be able to draw relevant conclusions on the potential use of nematic or other LC phases in specific applications. With this in mind, over the last years, our group has developed a stochastic method that can qualitatively and quantitatively mimic the Brownian motion of colloids as obtained by BD simulations  \cite{patti2012brownian, cuetos2015equivalence, corbett2018dynamic, daza2020dynamic}. This method, referred to as Dynamic Monte Carlo (DMC), has become an established simulation technique not only for the  study of the dynamics of biaxial particles, such as cuboids and curved rods \cite{chiappini2020, cuetos2020dynamics}, but also for the dynamics of uniaxial particles, like rods, for which soft potentials are indeed available \cite{levobka2019, chiappini_2020_2}. 

With regards to the equilibrium dynamics of cuboids, we found that the system long-time relaxation dramatically depends on particle anisotropy, being slower at the self-dual shape, the geometry that would preferentially stabilise biaxial nematics \cite{cuetos2020dynamics}. By definition, the self-dual shape is an intermediate geometry between prolate and oblate, where length ($L$), width ($W$) and thickness ($T$) are such that $W = \sqrt{LT}$. Our simulations also confirmed the occurrence of a Fickian and Gaussian dynamics at both short and long times, thus providing an alternative picture to the claimed universality of Fickian yet non-Gaussian dynamics in soft-matter systems \cite{morillo2018}. For its potential impact in nanotechnology, equally intriguing is the out-of-equilibrium dynamics of cuboids, especially because it can spark phase switching and new material properties. In general, the reorientation dynamics of biaxial particles induced by an external stimulus has received very limited attention. At the molecular scale, Lee and co-workers studied the reorientation dynamics of $\rm N_{B}$ phases of bent-core mesogens and measured primary and secondary axis switching, finding the latter either 3 or 100 times faster than the former depending on the mesogen \cite{lee2007dynamics}. Although this study was met with some scepticism \cite{stannarius2008comment}, Zannoni and co-workers later on performed MD simulations on $\rm N_{B}$ phases of biaxial Gay-Berne ellipsoids and confirmed that the rotation of minor axes is indeed faster, although only up to one order of magnitude, than the rotation of the main axis, both in the bulk \cite{berardi2008} and under confinement \cite{ricci2015field}. Following our recent findings on the field-induced stability of the $\rm N_B$ phase \cite{cuetos2019biaxial}, here we explore the field-induced dynamics of switching from uniaxial to biaxial nematics of colloidal cuboids, with special interest in the particle reorientation dynamics and associated response time. More specifically, we are interested to study the kinetics of reorientation of LCs transitioning between two different nematic textures, namely an $\rm N_{U}$ $\rightarrow$ $\rm N_{B}$ transition under an external field, and an $\rm N_{B}$ $\rightarrow$ $\rm N_{U}$ relaxation when the field is switched off, and estimate the associated response times. To gain an insight into the impact of particle anisotropy on the dynamics of phase switching, we  consider monodisperse systems of prolate, oblate and self-dual shaped cuboids. Their ability of reorienting under the effect of an external field is assessed by employing a DMC algorithm specifically designed to track the dynamics of out-of-equilibrium colloidal systems \cite{corbett2018dynamic}.

This paper is organised as follows. We first introduce the methodology to simulate our systems and characterise the dynamics. We then discuss the results of our simulations by analysing the effect of particle anisotropy on the out-of-equilibrium dynamics in $\rm N_{U}$ $\rightarrow$ $\rm N_{B}$ and $\rm N_{B}$ $\rightarrow$ $\rm N_{U}$ switching before finally drawing our conclusions.

\section{Model and Simulation Methodology}

\begin{figure} [h!]
	\includegraphics[width=1.00\linewidth,height=0.39\textheight]{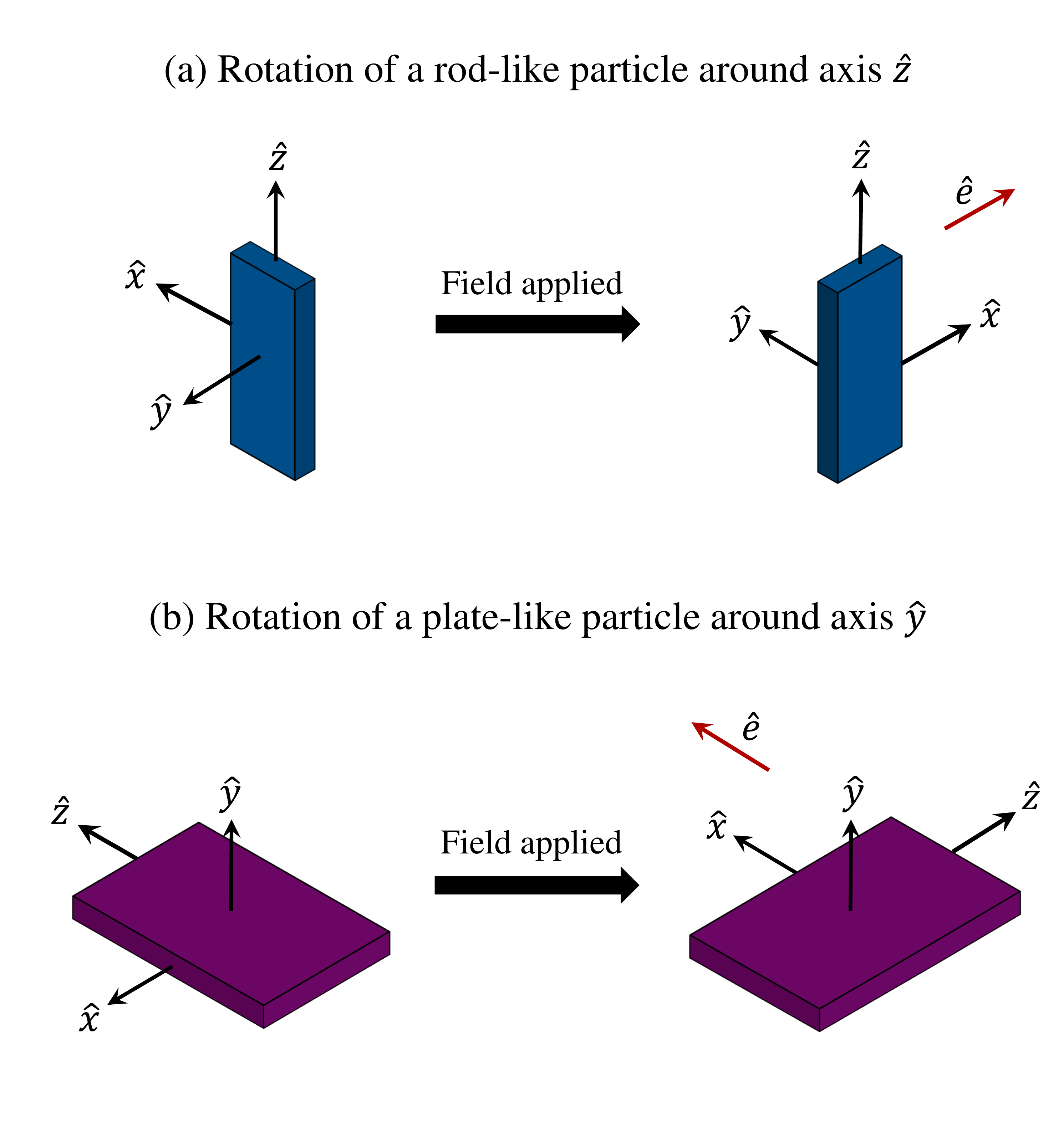}
	\caption{Reorientation of a prolate (a) and oblate (b) HBP due to external field $\rm \hat{\textbf{e}}$ coupled to the particle intermediate axis $\rm \hat{\textbf{x}}$. The $\rm N_{U}$ phases of self-dual shaped particles in this work always have a director along $\hat{\textbf{n}}$, and therefore rotate around $\hat{\textbf{z}}$ (similar to rod-like particles) when an external field is applied.}
	\label{fig:ext_field}
\end{figure}

We modelled monodisperse colloidal cuboids as hard board-like particles (HBPs) constrained in a cubic box with periodic boundaries. The behaviour of hard-core systems is basically determined by the packing fraction, which is given by:

\begin{equation}
    \eta \equiv \frac{Nv_{o}}{V}
\end{equation}

\noindent where $N$ is the number of particles, $v_{o}$ the volume of an individual HBP and $V$ the volume of the simulation box. The particle thickness, $T$, is set as the system unit length. Consequently, particle length and width are given in units of $T$, and read $L^* \equiv L/T$ and $W^* \equiv W/T$, respectively. In particular, $L^*=12$ for all systems studied, while $W^*$ assumed values between 1 (rod-like HBPs) and 12 (plate-like HBPs) and included $W^*=\sqrt{L^*}$ at the self-dual shape. Similar to our previous work \cite{cuetos2019biaxial}, we apply an external field that promotes alignment of the particle intermediate axis, defined by:

\begin{equation}
    U_{\rm ext} = -\frac{\varepsilon_{f}}{2} \Big[3 \cdot (\rm \hat{\textbf{x}}_{i} \cdot \rm \hat{\textbf{e}})^{2} -1 \Big]
\end{equation}

\noindent where $\varepsilon_f$ is the field strength, $\rm \hat{\textbf{x}}$ is the unit vector associated with the width of particle $i$, while $\rm \hat{\textbf{e}}$ is the field direction. We have set the reduced field strength, $\varepsilon^*_f \equiv \varepsilon_f \beta=3$, with $\beta$ the inverse temperature, which provides a measure of the relative strength of the field applied with respect to thermal energy. The unit vectors $\rm \hat{\textbf{y}}$ and $\rm \hat{\textbf{z}}$ are associated with thickness and length, respectively. Similar to our previous work, we remind that this external field model does not intend to mimic a real electric or magnetic field, but rather, we focus on its effect to reorient particles. The orientation of the particle unit vectors before and after application of the field is schematically displayed in Fig.\,\ref{fig:ext_field}.

The focus of this work is on the reorientation dynamics of HBPs in an external field. We first performed standard Monte Carlo (MC) simulations in the canonical ensemble in a cubic box containing $N = 2000$ HBPs to equilibrate $\rm N_{U}^{+}$ and $\rm N_{U}^{-}$ phases at $\eta = 0.34$, where $\rm N_{U}^{+}$ and $\rm N_{U}^{-}$ refer, respectively to prolate and oblate nematic LCs. At this packing fraction, the $\rm N_{U}$ phases (either prolate or oblate) are stable across all anisotropies \cite{cuetos2017phase}. Each MC cycle consists of $N$ attempts to displace and/or rotate HBPs, which are accepted if no overlaps are detected. To determine the occurrence of overlaps between pairs of HBPs, we implemented the separating axes theorem by Gottschalk \textit{et al.} \cite{gottschalk1996obbtree}, adapted by John and Escobedo to study tetragonal parallelepipeds \cite{john2005phase, john2008phase}. To quantify the system long-range orientational order, we calculated the nematic order parameter and director associated to each particle axis. To this end, we performed the diagonalisation of a second-rank symmetric tensor of the form:

\begin{equation}
\textbf{Q}^{\lambda \lambda} = \frac{1}{2N} \Biggl\langle \sum_{i=1}^{N} \big(3\hat{\lambda}_{i} \cdot \hat{\lambda}_{i} - \textbf{I}\big) \Biggr\rangle
\end{equation}

\noindent where $\hat{\lambda}_i$ = $\hat{\textbf{x}}$, $\hat{\textbf{y}}$, $\hat{\textbf{z}}$ are unit vectors aligned with \textit{W}, \textit{T} and \textit{L} respectively, while \textbf{I} is the identity tensor. The diagonalisation of $\textbf{Q}^{\lambda \lambda}$ results in three eigenvalues ($S_{2,W}$, $S_{2,T}$, $S_{2,L}$) and their corresponding eigenvectors ($\rm \hat{\textbf{m}}$, $\rm \hat{\textbf{l}}$, $\rm \hat{\textbf{n}}$). The nematic director of an $\rm N_U$ phase is the eigenvector with the largest eigenvalue. For instance, if the largest positive eigenvalue is $S_{2,L}$, then the nematic director is $\hat{\textbf{n}}$, indicating high degree of orientational order along the $\hat{\textbf{z}}$ axis of the particles. The biaxial order parameters can also be evaluated using the same symmetric tensor. For example, the biaxial order parameter that quantifies the fluctuations of particles' axes $\hat{\textbf{x}}$ and $\hat{\textbf{y}}$ respectively along the directors $\hat{\textbf{m}}$ and $\rm \hat{\textbf{l}}$ reads \cite{cuetos2017phase}

\begin{equation}
B_{2,L} = \frac{1}{3} \; (\hat{\bf m} \cdot \textbf{Q}^{xx} \cdot \hat{\bf m} + \hat{\bf l} \cdot \textbf{Q}^{yy} \cdot \hat{\bf l} - \hat{\bf m} \cdot \textbf{Q}^{yy} \cdot \hat{\bf m} - \hat{\bf l} \cdot \textbf{Q}^{xx} \cdot \hat{\bf l})
\end{equation}

\noindent The values of $B_{2,W}$ and  $B_{2,T}$ can be calculated using similar expressions. Following the definition introduced in our former work, a phase is considered to be biaxial if $B_{2} \geq 0.35$, although weak biaxial phases can already be observed for $B_{2} \geq 0.20$ \cite{cuetos2019biaxial, rafael2020self}. We monitor the evolution of uniaxial and biaxial order parameters until they have plateaued and fluctuate in a bounded range. The equilibrated configurations are then used for external field application in DMC simulations.

To study the dynamics, we performed DMC simulations in the canonical ensemble. Because our goal is producing realistic time trajectories, unphysical moves, such as cluster moves, swaps, jumps and changes in box dimension (which would result in centres of mass rescaling) are not implemented. The position of the particle centre of mass is updated by decoupling the displacement $\delta \textbf{r}_{i}$ into three contributions, with $\delta \textbf{r}_{i} = X_{W}\hat{\textbf{x}} + X_{T}\hat{\textbf{y}} + X_{L}\hat{\textbf{z}}$. Rotational moves are performed by three consecutive reorientations around $\hat{z}$, $\hat{x}$ and $\hat{y}$, with maximum rotations of $Y_{L}$, $Y_{W}$ and $Y_{T}$, respectively. The extent of particle displacement and rotation are picked from uniform distributions that depend on the particle translational, $D^{tra}_{\alpha,i}$, and rotational, $D^{rot}_{\alpha,i}$, diffusion coefficients at infinite dilution, with $\alpha={L,W,T}$. Maximum displacements and rotations are given by:

\begin{equation}
    \abs{X_{\alpha}} \leq \sqrt{2D^{tra}_{\alpha,i}\delta t_{MC}}
\end{equation}

\begin{equation}
    \abs{Y_{\alpha}} \leq \sqrt{2D^{rot}_{\alpha,i}\delta t_{MC}}
\end{equation}

\noindent where $\delta t_{MC}$ is the DMC timescale for one cycle, and is set to $\delta t_{MC} = 10^{-2} \tau$ for all simulations, with $\tau$ the time unit. The coefficients $D^{tra}_{\alpha,i}$ and $D^{rot}_{\alpha,i}$ have been estimated by using the open-source software HYDRO++ \cite{carrasco1999hydrodynamic,garcia2007improved}. The interested reader is referred to our previous work\,\cite{cuetos2020dynamics} for the specific values of these translational and rotational diffusivities in units of $D_{o} \equiv T^{2}\tau^{-1}$ and $D_{r} \equiv$ rad$^{2}\tau^{-1}$ respectively. For a monodisperse out-of-equilibrium system, the Brownian dynamics timescale, $\delta t_{BD}$, can be obtained by rescaling the MC time scale as follows

\begin{equation}
    \delta t_{BD} = \frac{\mathcal{A}_{c}}{3} \delta t_{MC}
\end{equation}

\noindent where $\mathcal{A}_{c}$ is the time-dependent acceptance rate calculated at the $c$th MC cycle over the transitory unsteady state \cite{corbett2018dynamic}. We determine $\mathcal{A}_{c}$ by performing an MC cycle at a fixed $\delta t_{MC}$ and integrating the above equation numerically:

\begin{equation}
  t_{\text{BD}}(\mathcal{C}_{\text{MC}}) = \delta t_{\text{MC}} \sum_{c=0}^{\mathcal{C}_{\text{MC}}} \frac{\mathcal{A}_{c}}{3}
\end{equation}

\noindent where $t_{\rm BD}(\mathcal{C}_{MC})$ is the Brownian time after $\rm \mathcal{C}_{MC}$ MC cycles. It should be noted that the DMC method does not consider solvent mediated hydrodynamic interactions, which are expected to become especially relevant at strong external fields or large packing fractions.

\begin{figure*} [htbp!]
	\centering
	\includegraphics[width=0.97\linewidth,height=0.258\textheight]{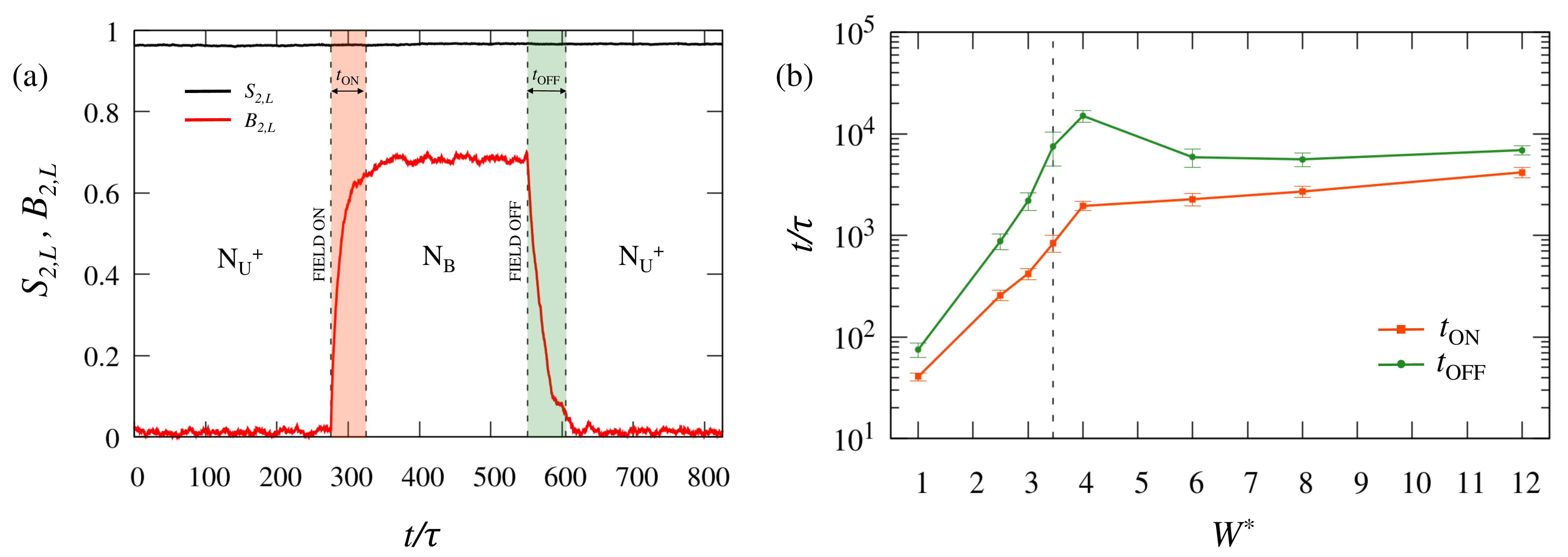}
	\caption{(a) Uniaxial and biaxial order parameters of a system with HBPs of $W^{*} = 1.0$ undergoing equilibration with an external field with $\varepsilon^*_f = 3$. The field is switched on at $t/\tau \approx 280$ and switched off at $t/\tau \approx 550$. The shaded orange area corresponds to $t_{\rm ON}$, while the green shaded area corresponds to $t_{\rm OFF}$. (b) Response times ($t_{\rm ON}$ and $t_{\rm OFF}$) as a function of $W^{*}$ with $\varepsilon^*_f = 3$. The dashed vertical line in (b) represents the self-dual shape that separates the prolate and oblate geometries.}
	\label{fig:Resp_times}
\end{figure*}

To characterise the dynamics, we estimated (\textit{i}) the response times, (\textit{ii}) the mean square angular displacement (MSAD), and (\textit{iii}) the angular self-part of the van-Hove function (s-VHF). We refer to the field-on ($t_{\rm ON}$) and field-off ($t_{\rm OFF}$) response times as the time taken for the biaxial order parameter to reach, respectively, 95$\%$ (field-on) and 105$\%$ (field-off) of its equilibrium value. In particular, when a field is applied to an $\rm N_{U}^{+}$ phase, $t_{\rm ON}$ is the time taken for $B_{2,L}$ to reach 95$\%$ of its equilibrium value in the field-on steady state. A schematic illustration of how we performed this evaluation is reported in Fig.\,\ref{fig:Resp_times}(a). Both sets of response times have been calculated from an average over 50 independent trajectories per system. Approximately 2$\%$ of these trajectories have given response times that were very different from those generally observed. Since these anomalies tend to distort averages and give misleadingly large error bars, we have considered them as outliers and excluded them from the average. To this end, we employed the Modified Z-score method, a multiple outlier rejection technique to identify statistical anomalies \cite{iglewicz1993detect}. In particular, the Modified Z-score, $M_{j}$, is given by the expression:

\begin{equation}
    M_{j} = \frac{0.6745 \times \big(t_{j} - \Bar{t}\big)}{MAD}
\end{equation}

\noindent where $t_{j}$ is the response time of trajectory $j$, $\Bar{t}$ is the median response time of the 50 trajectories and $MAD$ stands for median absolute deviation. An observable is considered an outlier only if $M_{j} > 3.5$ \cite{iglewicz1993detect}.

The MSAD provides the ensemble average of the particle angular displacements over time. To compute the MSAD, we employ a definition of an unbounded MSAD akin to the translational mean square displacement. To this end, we first introduce the definition of a rotational displacement vector which takes the form \cite{mazza2006relation,mazza2007connection}:

\begin{equation} \label{varphi}
    \overrightarrow{\varphi}(t) = \int_{0}^{t} \delta\overrightarrow{\varphi}(t') dt'
\end{equation}

\noindent where $ \delta\overrightarrow{\varphi}(t')$ is a vector with direction ${\hat{\lambda}_{i}}(t')  \times  \hat{\lambda}_{i}(t'+dt')$ and magnitude $| \delta\overrightarrow{\varphi}(t')|= \cos^{-1}[{\hat{\lambda}_{i}}(t') \cdot \hat{\lambda}_{i}(t'+dt')]$. From this, we can define the MSAD, which is mathematically expressed as:

\begin{equation}
    \langle \varphi^{2} (t)\rangle = \frac{1}{N} \Bigg\langle \sum\limits_{i=1}^N |{\overrightarrow{\varphi}}_{i}(t) - {\overrightarrow{\varphi}}_{i}(0)|^{2} \Bigg\rangle 
\end{equation}

\noindent where $\overrightarrow{\varphi}_{i}$ is the rotational displacement vector of particle $i$ defined in Eq.\,\ref{varphi}. Angular brackets denote average over different trajectories. Finally, the so-defined rotational displacements are employed to compute the angular s-VHF \cite{mazza2006relation,mazza2007connection}:

\begin{equation}
    G(\varphi,t) = \frac{1}{N} \Biggl\langle \sum\limits_{i=1}^N \delta(\varphi - |\overrightarrow{\varphi}_{i}(t+t_{0}) - \overrightarrow{\varphi}_{i}(t_{0})|)  \Biggr\rangle
\end{equation}

\noindent where the symbol $\delta$ is the Dirac delta function. Basically, $G(\varphi,t)$ provides the probability distribution of angular displacements of particles within a time $t+t_{0}$ given their position at time $t_{0}$.


\section{Results}

\begin{figure*}[htbp!]
	\includegraphics[width=0.9\linewidth]{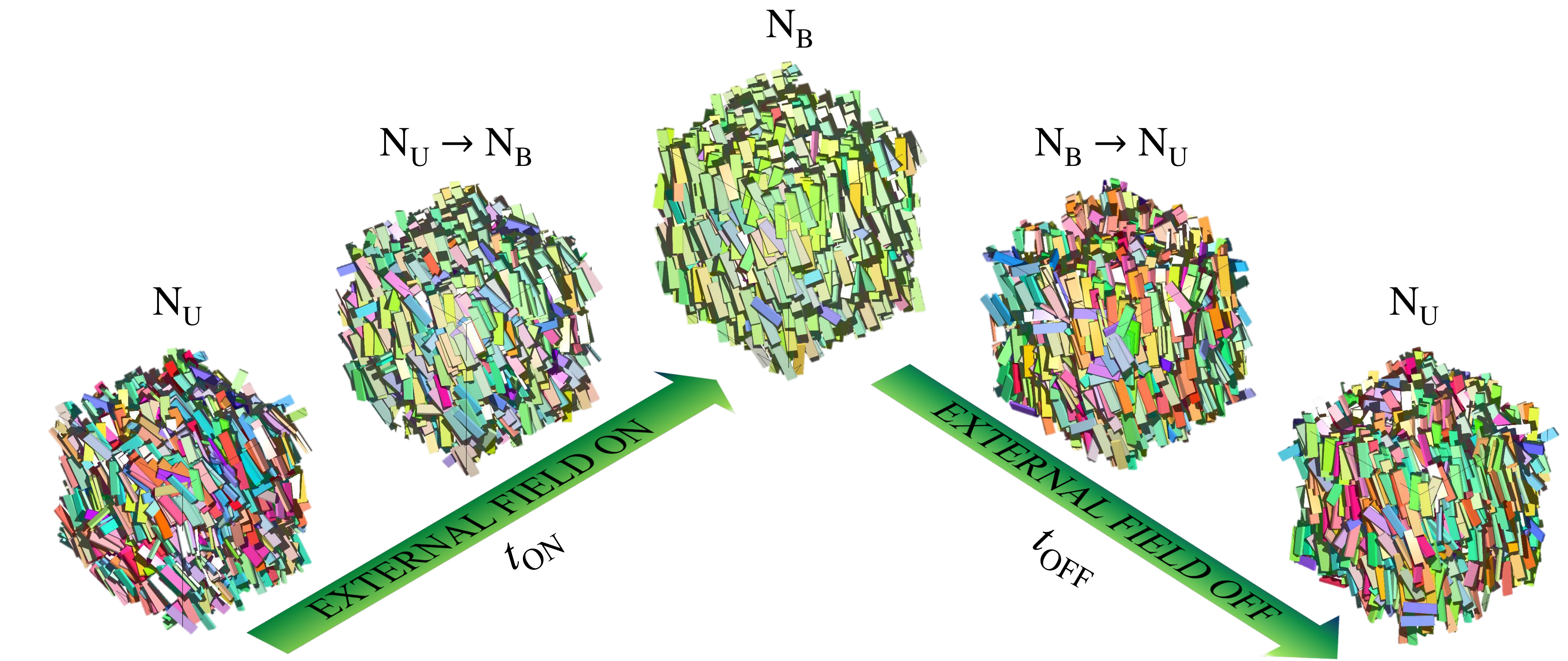}
	\caption{Schematic illustration of a field-induced $\rm N_{U}$ $\rightarrow$ $\rm N_{B}$ and a free $\rm N_{B}$ $\rightarrow$ $\rm N_{U}$ switching. Different colours represent different orientations. Readers interested on the detailed structure of the $\rm N_{U}$ and $\rm N_{B}$ phases are referred to our past works for snapshots with higher resolution \cite{cuetos2019biaxial,rafael2020self}.}
	\label{fig:switching}
\end{figure*}

Upon application of a sufficiently strong external field, an $\rm N_U$ phase can be transformed into an $\rm N_{B}$ phase \cite{cuetos2019biaxial}. This transformation is not permanent, and, when the field is removed, uniaxiality is restored. The time taken by the particles to reorient along the field director measures the system's ability of switching to a more ordered configuration. Vice-versa, when the field is removed, the particles are left free to rotate and the system recovers its original uniaxial state. A schematic illustration of both transitory states is given in Fig.\,\ref{fig:switching}. We have measured the response time associated to both $\rm N_{U}$ $\rightarrow$ $\rm N_{B}$ and $\rm N_{B}$ $\rightarrow$ $\rm N_{U}$ transitions upon application of the field $U_{\rm ext}$ with $\varepsilon^*_f=3$. The effect of changing field intensity between $\varepsilon^*_f=1.5$ and 3 on the $\rm N_{U}$ $\rightarrow$ $\rm N_{B}$ response time has also been assessed and is available, for the interested reader, in Appendix A. The resulting response times, $t_{\rm ON}$ and $t_{\rm OFF}$, are reported in Fig\,\ref{fig:Resp_times}(b). Since the formation of a field-induced $\rm N_{B}$ phase is dependent on the alignment of the particle intermediate axis $\hat{x}$ with the external field, the discussion that follows is relative to this axis, unless otherwise stated. To start with, we notice that $t_{\rm ON} < t_{\rm OFF}$ across the complete set of anisotropies (see Fig.\,\ref{fig:Resp_times}(b)). In other words, at a given particle width, the $\rm N_{U}$ $\rightarrow$ $\rm N_{B}$ switching is faster than the $\rm N_{B}$ $\rightarrow$ $\rm N_{U}$ switching. To understand the origin of this behaviour, we calculated the MSADs of our systems and compared the field-on and field-off profiles for each anisotropy.

\begin{figure}[htbp!]
	\centering
	\includegraphics[width=0.95\linewidth,height=0.5\textheight]{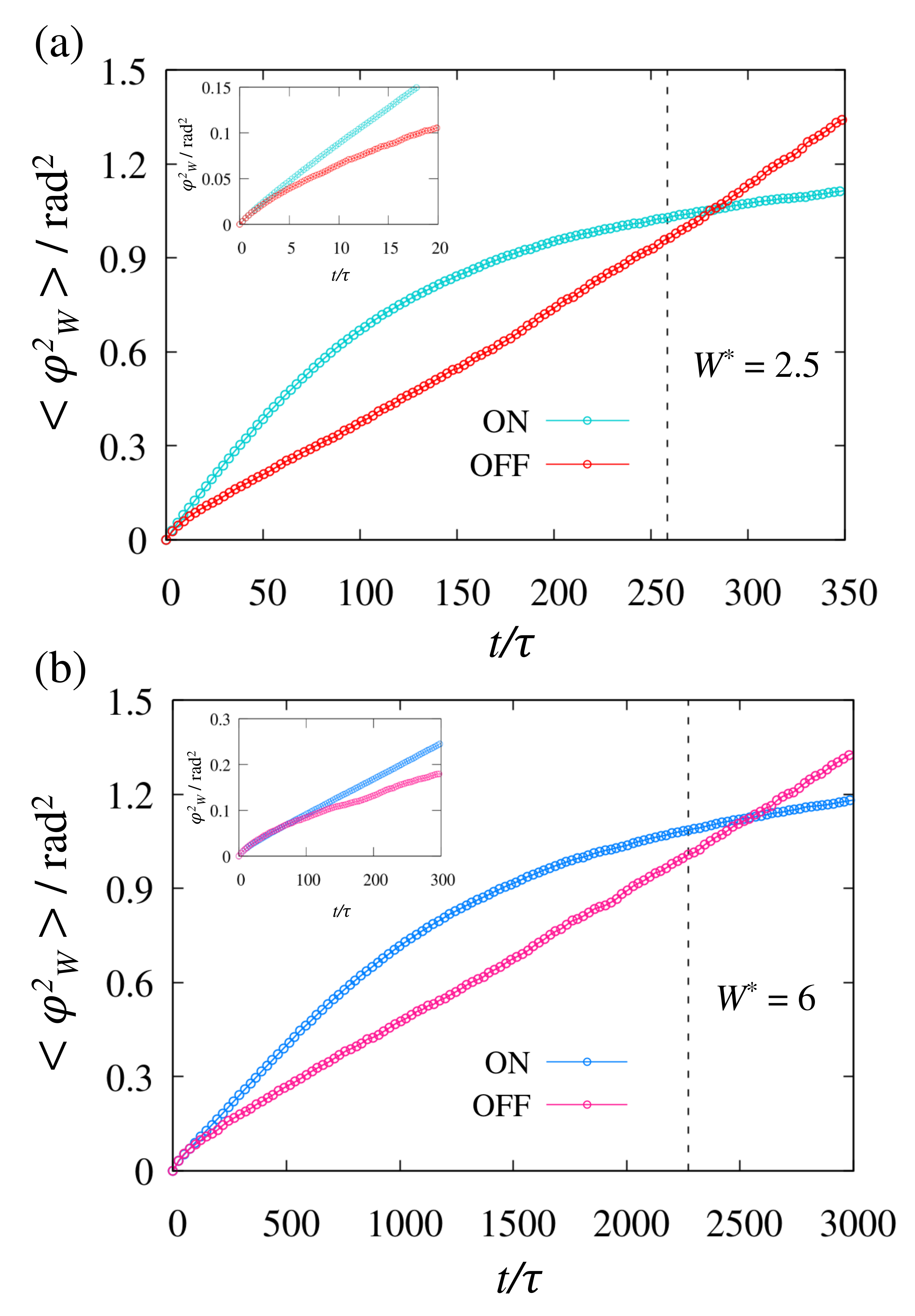}
	\caption{MSAD in field-on and field-off scenarios of a system of HBPs with reduced width (a) $W^{*} = 2.5$ and (b) $W^{*} = 6$. The field-on simulations apply an external field of strength $\varepsilon^*_{f} = 3$. The dashed vertical lines indicate $t_{\rm ON}$ of each systems ($t_{\rm OFF}$ is out of scale and not shown). The insets in (a) and (b) show the MSAD at shorter timescales.}
	\label{fig:MSAD_1}
\end{figure}

The MSAD of systems with $W^{*} = 2.5$ and 6 are shown, respectively, in the top and bottom frames of Fig.\,\ref{fig:MSAD_1}. At very short times, the field-on and field-off MSADs are very similar to each other, with the former becoming larger immediately after and up to relatively long time scales. Over this period of time, field-induced rotation is faster than free rotation. However, on time scales comparable to $t_{\rm ON}$, a crossover between the two MSADs is observed. On these time scales and beyond, free rotation grows significantly much faster with time than field-induced rotation. We therefore conclude that the presence of the external field accelerates the system orientational dynamics by forcing the reorientation of the particle $\hat{x}$ axis along the field director. As more and more HBPs are oriented, the field-on MSAD grows less and less with time and would eventually saturate to a plateau if the field strength was sufficiently large to offset and overcome thermal forces. By contrast, the field-off MSAD practically shows the same behaviour with time over the full time scale, as expected in free rotational diffusion. As for the effect of anisotropy on the response time, we first discuss the case of the field-induced uniaxial-to-biaxial transitory state. 

\begin{figure*}[htbp!]
	\centering
	\includegraphics[width=1.00\linewidth,height=0.365\textheight]{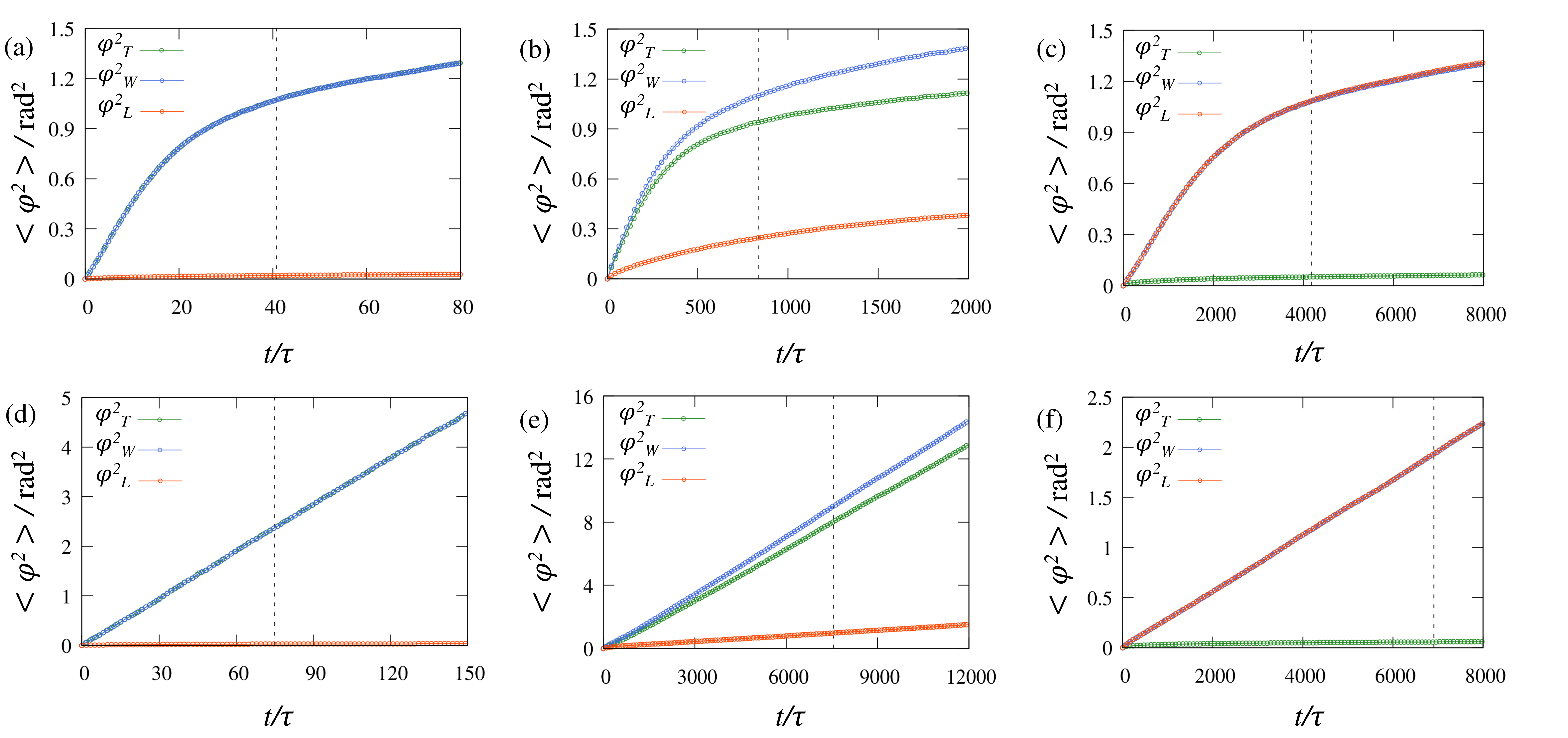}
	\caption{MSAD of the $\hat{x}$, $\hat{y}$ and $\hat{z}$ axes of HBPs for systems undergoing $\rm N_{U} \rightarrow N_{B}$ transition with $\varepsilon^*_f = 3$ for (a) prolate; (b) self-dual and (c) oblate HBPs, and $\rm N_{B} \rightarrow N_{U}$ transition when the field is switched off for a (d) prolate; (e) self-dual and (f) oblate HBPs. The dashed vertical lines in each figure represent the $t_{\rm ON}$ for (a)-(c) and $t_{\rm OFF}$ for (d)-(f).}
	\label{fig:MSAD_2}
\end{figure*}

In Fig. \ref{fig:Resp_times}(b), $t_{\rm ON}$ increases with $W^*$, implying that the reorientation is slower for oblate than for prolate particles. More specifically, for rod-like HBPs ($W^{*} = 1$), we observe a rapid switching with $t_{\rm ON}/\tau \approx 41$, whereas for plate-like HBPs ($W^{*} = 12$), it is significantly slower, with $t_{\rm ON}/\tau \approx 4200$. Consequently, making HBPs more oblate leads to a slower field-induced $\rm N_{U} \rightarrow N_{B}$ transition. To confirm these preliminary tendencies, we compare the MSADs of the field-on regimes of each anisotropy along the three axes. The top frames of Fig.\,\ref{fig:MSAD_2} display the field-on MSADs of systems containing HBPs with $W^{*} = 1$, 3.46 and 12. We notice that the MSAD of the particle axis oriented as the nematic director of the original $\rm N_U$ phase is the smallest across all the geometries. More specifically, the MSAD of rod-like particles in Fig.\,\ref{fig:MSAD_2}(a) exhibits a strong rotational coupling between $\hat{x}$ and $\hat{y}$ particle axes, while $\hat{z}$ is practically unaffected by the application of the field. Such a strong angular correlation between $\hat{x}$ and $\hat{y}$, with $\langle \varphi^2_W \rangle= \langle \varphi^2_T \rangle$ over time, is due to the square cross-sectional area of this specific set of HBPs, where $W=T$. For similar reasons, plate-like HBPs with $W=L$ exhibit strong rotational correlations between their axes $\hat{x}$ and $\hat{z}$, with $\langle \varphi^2_W \rangle= \langle \varphi^2_L \rangle$ (see Fig.\,\ref{fig:MSAD_2}(c)), while $\langle \varphi^2_T \rangle$, slightly increasing over time for mere thermal fluctuations, remains very small, practically insensible to the external field. In systems of self-dual shaped HBPs ($W^{*} = 3.46$), we observe that the MSADs of $W$ and $T$ are initially coupled, but  then diverge over time. This behaviour is observed for all anisotropies that are not perfectly rod-like or plate-like and agrees very well with the tendencies reported in our recent work on the equilibrium dynamics of HBPs \cite{cuetos2020dynamics}.

When analysing the field-on MSADs of the particle axes perpendicular to the original nematic director, we also notice an initially linear, rather steep dependence on time, followed by an intermediate non-linear behaviour and subsequently by a second linear regime at times comparable to $t_{\rm ON}$. Such a long-time linear regime suggests that HBPs' angular displacements are gradually reducing, due to the system approaching a new equilibrium state. Under these conditions, further rotations of the particle intermediate axis $\hat{x}$, which is already aligned with the field, are suppressed, and only small angular fluctuations are detected. At much larger field strengths, with thermal fluctuations completely inhibited, we expect this second linear regime to plateau at long times. In agreement with Fig.\,2(b), we also observe that systems with rod-like HBPs ($W^{*} = 1$) only take $t/\tau \approx 17$ to reach $<\varphi^{2}_{W}> = 0.6$ rad$^2$, whereas systems with self-dual shaped ($W^{*} = 3.46$) or plate-like ($W^{*} = 12$) HBPs take, respectively, $t/\tau \approx 250$ and $t/\tau \approx 1600$ to achieve the same MSAD value. This suggests that prolate HBPs tend to reorient significantly faster when an external field is applied, leading to a relatively rapid equilibration. 

\begin{figure} [htbp!]
	\centering
	\includegraphics[width=1.0\linewidth,height=0.53\textheight]{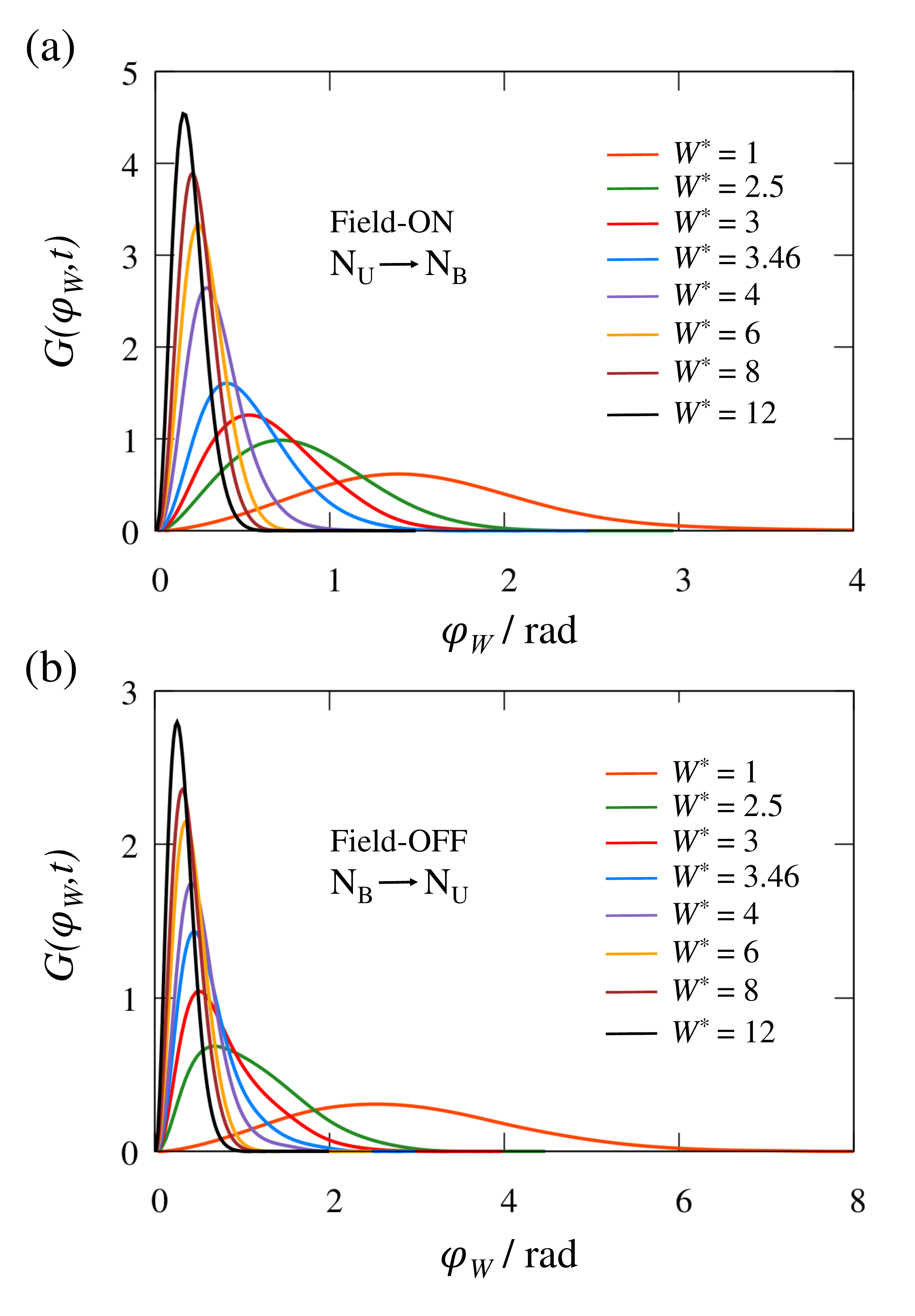}
	\caption{(a) Angular s-VHFs of (a) field-on reorientation at time $t/\tau = 35$ and (b) field-off reorientation at time $t/\tau = 100$.}
	\label{fig:s_VHF}
\end{figure}

To gain a better insight into the dynamics of reorientation during this first transitory unsteady state, we calculated the s-VHFs of all anisotropies at $t/\tau = 35$, corresponding to a time when all field-on cases are still undergoing equilibration. As the MSAD for field-off scenarios are linear, we arbitrarily picked $t/\tau = 100$ to show the field-off s-VHFs. The s-VHFs shown in Fig.\,\ref{fig:s_VHF} refer to the intermediate axis and have been normalised such that $\int_{0}^{\infty} 4\pi\varphi^{2}G(\varphi,t) d\varphi = 1$. The first evident conclusion, confirming the results discussed so far, is that prolate HBPs rotate faster than oblate HBPs. This can be appreciated in Fig.\,\ref{fig:s_VHF}(a) by pinpointing the location of the peak of $G(\varphi_W,t)$, which indicates the most probable rotation achieved by particles of a given geometry at $t/\tau = 35$. In particular, the peak of $G(\varphi_{W=T},t)$ and $G(\varphi_{W=L},t)$ suggests that rod-like and plate-like particles have rotated, respectively, by $\varphi_{W} \approx 1.4$ rad and $\varphi_{W} \approx 0.2$ rad. By increasing particle width from $W^*=1$ to 12, the peak of the angular s-VHFs gradually displaces towards lower rotations. Not only does the particle anisotropy determine the location of the peak of these distributions at a given time, but also their broadness. In other words, the angular s-VHFs provide relevant information on the most probable rotation performed by HBPs and on the existence of HBPs that rotate faster or slower than the average. In particular, the presence of fast- and slow-responsive HBPs is evinced by the tails of the distributions in  Fig.\,\ref{fig:s_VHF}(a), especially broad at $W^*=1$ and then narrower and narrower up to $W^*=12$. Therefore, rod-like HBPs rotate relatively fast, but heterogeneously (broad $G(\varphi_W,t)$ peaked at large distances), whereas plate-like HBPs are significantly slower, but rotate much more homogeneously (narrow $G(\varphi_W,t)$ peaked at short distances). 

\begin{figure} [htbp!]
	\centering
	\includegraphics[width=1.0\linewidth,height=0.53\textheight]{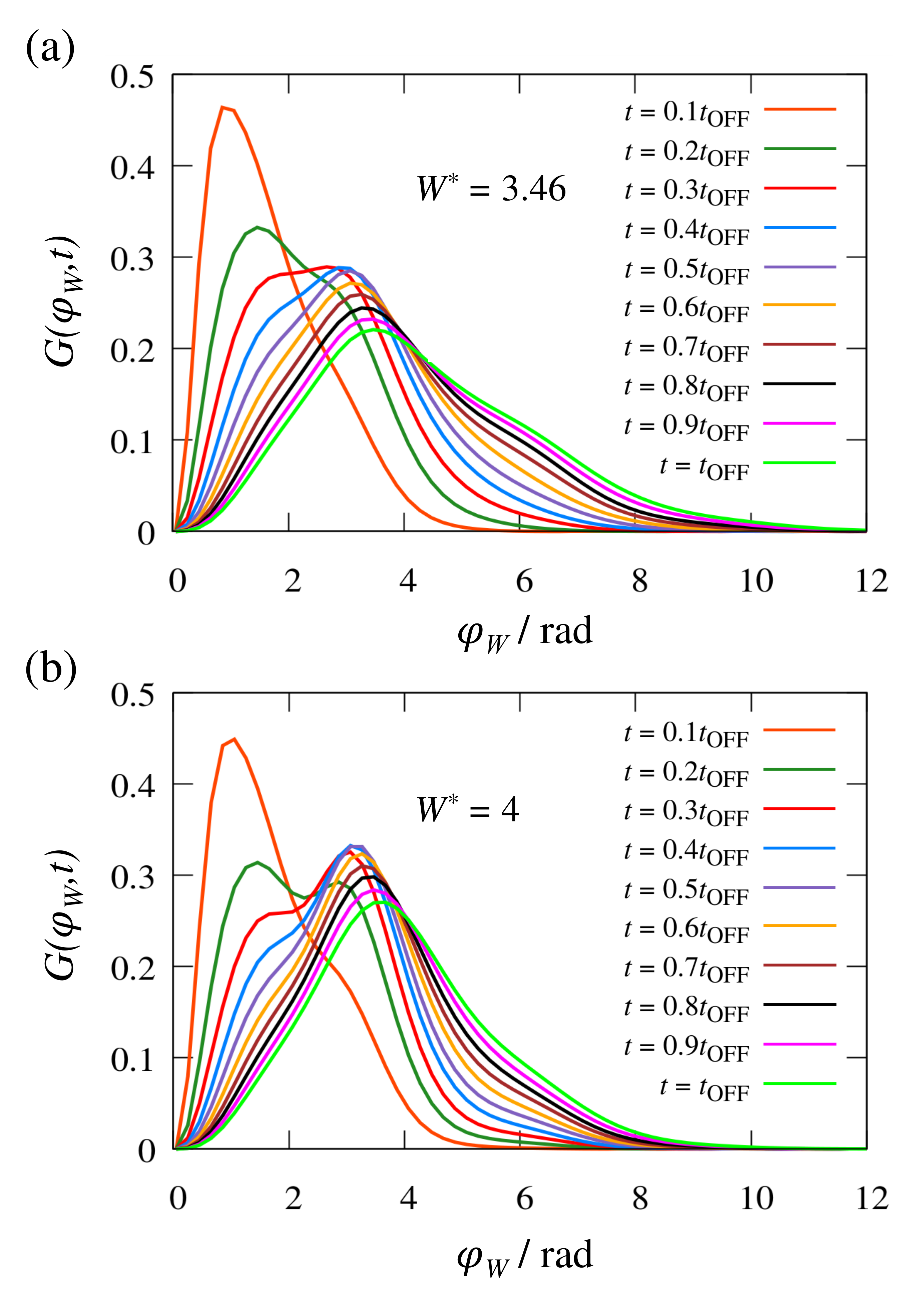}
	\caption{(a) Angular s-VHFs at various times, expressed as percentage of $t_{\rm OFF}$, in systems of HBPs with (a) $W^{*} = 3.46$ and (b) $W^{*} = 4$.}
	\label{fig:s_VHF_2}
\end{figure}

When the field is switched off, the system recovers its original uniaxial symmetry with the particles free to reorient under the mere effect of thermal fluctuations. The time $t_{\rm OFF}$ taken by this field-off reorientation to re-establish the $\rm N_U$ phase is again much shorter in nematics of prolate HBPs (see Fig.\,\ref{fig:Resp_times}(b)). In particular, $t_{\rm OFF}$ shows a tendency to increase with particle width up to $W^{*} = 4$, where $t_{\rm OFF}/\tau \approx 15 \cdot 10^3$. When HBPs acquire a modest oblate geometry, such as from $W^{*} = 6$ to $W^{*} = 8$, $t_{\rm OFF}$ drastically decreases to around $t_{\rm OFF}/\tau \approx 6 \cdot 10^3$, before increasing again at $W^{*} = 12$. We also notice that the free reorientation at $W^{*} = 3.46$ and 4 is particularly slower than that of other anisotropies and deserves an explanation. The MSADs in the field-off scenarios, shown in Fig.\,\ref{fig:MSAD_2}(d)-(f), exhibit a linear profile throughout the simulation due to the absence of an external field and decrease upon increasing $W^{*}$. This tendency is also detected in the field-off s-VHFs of Fig.\,\ref{fig:s_VHF}(b), where, similarly to the field-on case, the angular displacement decreases at increasing particle width. These elements would suggest a scenario where $t_{\rm OFF}$ increases and free rotation becomes slower upon increasing $W^{*}$. Because $t_{\rm OFF}$ is peaked for approximately self-dual shaped particles and then decreases (see Fig.\,\ref{fig:Resp_times}(b)), there must be an additional element contributing to the field-free reorientation from the $\rm N_B$ to the $\rm N_U$ phase. We believe that this element is related to the ability of self-dual shaped HBPs of retaining phase biaxiality when the field is switched off. In our recent work on the field-induced phase behaviour of HBPs, we found that the self-dual shape requires a surprisingly weak external field, compared to prolate and oblate geometries, to spark an $\rm N_{U} \rightarrow N_{B}$ transition \cite{cuetos2019biaxial}. In particular, the minimum field strength to stabilise $\rm N_{B}$ LCs was found to be $\varepsilon^*_f=0.1$ and 0.25 at $W^*=3.46$ and 4, respectively, and then increasing to $\varepsilon^*_f=0.5$ at $W^*=3$ and to $\varepsilon^*_f=1$ at $W^*=6$. Therefore, we believe that the field free $\rm N_{U} \rightarrow N_{B}$ transition at or very close to the self-dual shape is affected by a metastability of the $\rm N_{B}$ phase in off-field case. The underlying metastability allows the system to retain the induced biaxiality over a longer time when compared with nematics of oblate of prolate HBPs. In Fig.\,\ref{fig:s_VHF_2}, we show the evolution of the s-VHFs of $W^{*} = 3.46$ and $W^{*} = 4$ at different times up to $t_{\rm OFF}$. At short to intermediate time scales, after having switched the field off, these s-VHFs exhibit a double peak that suggests the presence of two populations of HBPs. Because these two populations rotate at sufficiently different rates, we can label them as  \textit{slow} and \textit{fast}. The first peak survives over a relatively long period of time, between 0.2$t_{\rm OFF}$ and 0.7$t_{\rm OFF}$, turning gradually into a shoulder that disappears at longer times. These peaks and subsequent shoulders are especially pronounced in the case of $W^{*} = 4$ (Fig.\,\ref{fig:s_VHF_2}(b)), explaining why $t_{\rm OFF}$ at $W^{*} = 4$ is significantly slower than $t_{\rm OFF}$ at $W^{*} = 3.46$. Double peaks and shoulders are not observed at other anisotropies or in field-on transitions (not shown here), indicating that these tendencies are especially relevant only in the field-off relaxation of self-dual shaped particles. In addition to their propensity towards biaxiality retention, HBPs with $W^*=4$ rotate more slowly than perfectly self-dual shaped particles, as shown in Fig.\,\ref{fig:s_VHF}(b) and in agreement with the tendencies observed in $\rm N_U$ phase in the absence of external fields \cite{cuetos2020dynamics}. The resulting large value of $t_{\rm OFF}$ is therefore determined by the interplay between the particle's ability to rotate and the system's tendency of retaining phase biaxiality. This interplay explains the non-monotonic trend of $t_{\rm OFF}$ with particle shape in Fig.\,\ref{fig:Resp_times}(b)) and provides, along with $t_{\rm ON}$, a useful guideline to select the most suitable particle anisotropy for the design of field-responsive nanomaterials.

\section{Conclusions}

In summary, by Dynamic Monte Carlo simulation, we studied the field-induced dynamics in uniaxial nematic LCs of colloidal HBPs. By forcing the particles to reorient around the nematic director, the external field induces an $\rm N_{U} \rightarrow N_{B}$ phase transition that takes the system to a new steady state. When the field is switched off, the biaxiality is gradually lost and the $\rm N_U$ phase is restored. The time taken for the system to reorient, also referred to as response time, strongly depends on the particle anisotropy. The response times in $\rm N_{U} \rightarrow N_{B}$ and $\rm N_{B} \rightarrow N_{U}$ switching were calculated and compared across all anisotropies studied. Despite being the optimal shape to promote phase biaxiality, the switching dynamics of self-dual shape HBPs is less satisfactory compared to prolate HBPs. In particular, rod-like HBPs with $W^{*} = 1$ exhibit the fastest reorientation times in both the field-on and field-off cases. The analysis of MSADs and s-VHFs show that the response time is a result of a trade-off between particle rotational diffusion and phase biaxiality retention, being both determined by shape anisotropy. Prolate HBPs were found to rotate faster than self-dual shaped or oblate HBPs, allowing rapid phase switching between the two nematic phases. Systems of HBPs with geometry equal or very close to the self-dual shape exhibit a particularly slow field-free reorientation, most likely due to relatively low field strength required to transform $\rm N_U$ into $\rm N_B$ phases and favouring the former in absence of an external field \cite{cuetos2019biaxial}. The ability of retaining biaxiality over a longer period of time is corroborated by the existence of a double peak in the angular s-VHFs of $\rm N_{B} \rightarrow N_{U}$ transition at short-to-intermediate time scales. This double peak suggests the existence of two populations of (quasi) self-dual shaped HBPs whose reorientation is not uniform and delays the system relaxation.  While prolate HBPs are especially field-responsive and exhibit a rapid field-free reorientation, when one analyses the distribution of their angular displacements over time, this appears to be very broad, with particles exhibiting a very heterogeneous ability of rotating. By contrast, oblate HBPs, while significantly less responsive, are characterised by a very narrow distribution of angular displacements. All these elements offer a fundamental understanding of the impact of shape anisotropy on the dynamics of uniaxial-to-biaxial switching and a guidance to formulate nanomaterials with specific switching dynamics for target applications.

\section{Acknowledgements}

EMR would like to thank the Malaysian Government Agency Majlis Amanah Rakyat for funding his PhD at the University of Manchester. AP and LT acknowledges the financial support from the Leverhulme Trust Research Project Grant RPG-2018-415. AC acknowledge the Spanish Ministerio de Ciencia, Innovación y Universidades and FEDER for funding (project PGC2018-097151-B-I00). EMR, LT, DC and AP acknowledge the assistance given by IT Services and the use of Computational Shared Facility at the University of Manchester. Finally, we thank Gerardo Campos-Villalobos (Utrecht University) for his assistance in generating the snapshots in Fig.\,\ref{fig:switching}.

\section{Data Availability}
The data that support the findings of this study are available from the corresponding author upon reasonable request.


\appendix

\section{Effect of Field Strength}

\begin{figure} [htbp!]
	\centering
	\includegraphics[width=0.92\linewidth,height=0.285\textheight]{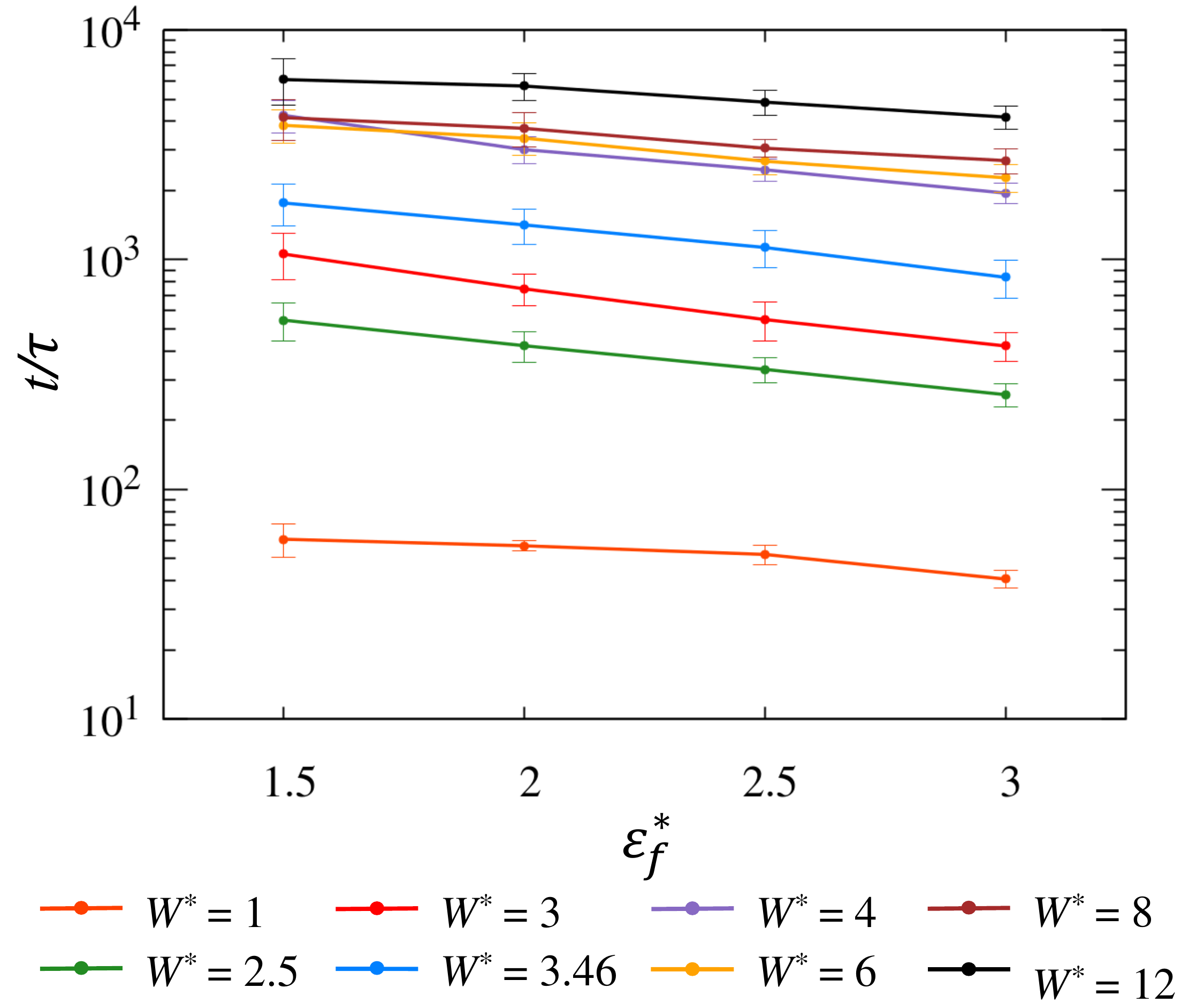}
	\caption{Changes in field-on response time, $t_{\rm ON}$ as a function of $\varepsilon^*_f$ across different anisotropies.}
	\label{fig:var_field}
\end{figure}

In this appendix, we briefly discuss the effect of altering field intensity on the field-on response times, $t_{\rm ON}$ of HBPs. Here, we report the response times for the field-on case for all anisotropies studied at field strengths from $\varepsilon^*_f = 1.5$ to $\varepsilon^*_f = 3$. These field intensities result in the formation of strong $\rm N_{B}$ phases with $B_{2} \geq 0.35$ \cite{cuetos2019biaxial}. The results are shown in Fig. \ref{fig:var_field}.

At constant $\varepsilon^*_{f}$, we observe that $t_{\rm ON}$ generally increases with $W^{*}$ and this increment is significant. For instance, at $\varepsilon^*_{f} = 2$, the response time increases by two orders of magnitude from $t_{\rm ON} \approx 57$  at $W^{*} = 1$ to $ t_{\rm ON} \approx 5700$ at $W^{*} = 12$. We conclude that prolate particles tend to rotate faster than oblate particles, regardless the field strength. Upon increasing field strength, the reorientation becomes faster as suggested by the gradual decrease of $t_{\rm ON}$ with $\varepsilon^*_{f}$. In addition, we note that the statistical errors in $t_{\rm ON}$ decrease with $\varepsilon^*_{f}$, most likely due to a stronger suppression of rotational fluctuations. This tendency is consistent with the works by Zannoni and co-workers \cite{berardi2008}. For rod-like HBPs ($W^{*} = 1$), increasing $\varepsilon_{f}$ does not significantly affect $ t_{\rm ON}$, probably because the reorientation capability of these HBPs is very close to its saturation value.

\section{Response Times}

\begin{table*} [htbp!]
	\begingroup
	\setlength{\tabcolsep}{14pt} 
	\renewcommand{\arraystretch}{1.1} 
\begin{tabular}{lccccc}
\hline
\hline
 $W^{*}$ & \multicolumn{4}{c}{$t_{\rm ON}$} & {$t_{\rm OFF}$} \\
\cline{2-5}
& $\varepsilon^*_f = 1.5 $ & $\varepsilon^*_f = 2 $ & $\varepsilon^*_{f} = 2.5 $ & $\varepsilon^*_{f} = 3 $ & $\varepsilon^*_{f} = 3 $\\
\hline
1     & $60.7 \pm 10.0$   & $56.8 \pm 2.9$   & $52.1 \pm 5.2$ & 40.7 $\pm$ 3.6 & 75.0 $\pm$ 12.4 \\ 
2.5   & $545.2 \pm 100.9$  & $422.0 \pm 64.4$   & $332.9 \pm 40.7$ & 258.3 $\pm$ 29.4 & 878.6 $\pm$ 160.1 \\ 
3     & $1057.5 \pm 243.0$  & $746.4 \pm 118.2$   & $548.4 \pm 106.8$ & 420.9 $\pm$ 59.2 & 2181.0 $\pm$ 434.8 \\
3.46  & $1764.9 \pm 365.6$  & $1416.6 \pm 247.5$  & $1130.0 \pm 205.6$ & 837.6 $\pm$ 160.0 & 7552.2 $\pm$ 2758.4\\
4     & $4247.8 \pm 683.3 $ & $3012.6 \pm 403.6$  & $2461.8 \pm 273.4$ & 1948.8 $\pm$ 200.5 & 15004.6 $\pm$ 2011.0\\ 
6     & $3844.7 \pm 641.8$  & $3380.6 \pm 546.0$  & $2681.7 \pm 345.7$ & 2272.2 $\pm$ 310.4 & 5975.5 $\pm$ 1200.2 \\
8     & $4151.4 \pm 846.8$  & $3728.3 \pm 636.6$  & $3055.4 \pm 272.5$ & 2700.0 $\pm$ 340.2 & 5602.0 $\pm$ 877.4\\ 
12    & $6084.4 \pm 1376.9$  & $5700.9 \pm 777.0$  & $4849.0 \pm 598.6$ & 4175.5 $\pm$ 487.5 & 6906.0 $\pm$ 733.8\\ 
\hline
\hline
\end{tabular}
	\endgroup
\caption{Table of response times for $t_{\rm ON}$ ($\varepsilon^*_{f} = 1.5$ to $3$) and $t_{\rm OFF}$ ($\varepsilon^*_{f} = 3)$ with associated statistical errors.}
\label{tab:response}
\end{table*}

In Table I, we report $t_{\rm ON}$ for $\varepsilon^*_{f}=1.5$, 2, 2.5 and 3, and $t_{\rm OFF}$ for $\varepsilon^*_{f} = 3$.

\bibliography{POF_production.bib}

\begin{thebibliography}{50}%
\makeatletter
\providecommand \@ifxundefined [1]{%
 \@ifx{#1\undefined}
}%
\providecommand \@ifnum [1]{%
 \ifnum #1\expandafter \@firstoftwo
 \else \expandafter \@secondoftwo
 \fi
}%
\providecommand \@ifx [1]{%
 \ifx #1\expandafter \@firstoftwo
 \else \expandafter \@secondoftwo
 \fi
}%
\providecommand \natexlab [1]{#1}%
\providecommand \enquote  [1]{``#1''}%
\providecommand \bibnamefont  [1]{#1}%
\providecommand \bibfnamefont [1]{#1}%
\providecommand \citenamefont [1]{#1}%
\providecommand \href@noop [0]{\@secondoftwo}%
\providecommand \href [0]{\begingroup \@sanitize@url \@href}%
\providecommand \@href[1]{\@@startlink{#1}\@@href}%
\providecommand \@@href[1]{\endgroup#1\@@endlink}%
\providecommand \@sanitize@url [0]{\catcode `\\12\catcode `\$12\catcode
  `\&12\catcode `\#12\catcode `\^12\catcode `\_12\catcode `\%12\relax}%
\providecommand \@@startlink[1]{}%
\providecommand \@@endlink[0]{}%
\providecommand \url  [0]{\begingroup\@sanitize@url \@url }%
\providecommand \@url [1]{\endgroup\@href {#1}{\urlprefix }}%
\providecommand \urlprefix  [0]{URL }%
\providecommand \Eprint [0]{\href }%
\providecommand \doibase [0]{http://dx.doi.org/}%
\providecommand \selectlanguage [0]{\@gobble}%
\providecommand \bibinfo  [0]{\@secondoftwo}%
\providecommand \bibfield  [0]{\@secondoftwo}%
\providecommand \translation [1]{[#1]}%
\providecommand \BibitemOpen [0]{}%
\providecommand \bibitemStop [0]{}%
\providecommand \bibitemNoStop [0]{.\EOS\space}%
\providecommand \EOS [0]{\spacefactor3000\relax}%
\providecommand \BibitemShut  [1]{\csname bibitem#1\endcsname}%
\let\auto@bib@innerbib\@empty
\bibitem [{\citenamefont {Mormann}\ \emph {et~al.}(2017)\citenamefont
  {Mormann}, \citenamefont {Hellwich}, \citenamefont {Chen},\ and\
  \citenamefont {Wilks}}]{mormann2017preferred}%
  \BibitemOpen
  \bibfield  {author} {\bibinfo {author} {\bibfnamefont {W.}~\bibnamefont
  {Mormann}}, \bibinfo {author} {\bibfnamefont {K.-H.}\ \bibnamefont
  {Hellwich}}, \bibinfo {author} {\bibfnamefont {J.}~\bibnamefont {Chen}}, \
  and\ \bibinfo {author} {\bibfnamefont {E.~S.}\ \bibnamefont {Wilks}},\
  }\href@noop {} {\bibfield  {journal} {\bibinfo  {journal} {Pure Appl. Chem.}\
  }\textbf {\bibinfo {volume} {89}},\ \bibinfo {pages} {1695} (\bibinfo {year}
  {2017})}\BibitemShut {NoStop}%
\bibitem [{\citenamefont {Brown}(2015)}]{brown_2015}%
  \BibitemOpen
  \bibfield  {author} {\bibinfo {author} {\bibfnamefont {R.}~\bibnamefont
  {Brown}},\ }\href {\doibase 10.1017/CBO9781107775473} {\emph {\bibinfo
  {title} {The Miscellaneous Botanical Works of Robert Brown}}},\ edited by\
  \bibinfo {editor} {\bibfnamefont {J.~J.}\ \bibnamefont {Bennett}},\ \bibinfo
  {series} {Cambridge Library Collection - Botany and Horticulture},
  Vol.~\bibinfo {volume} {1}\ (\bibinfo  {publisher} {Cambridge University
  Press},\ \bibinfo {year} {2015})\BibitemShut {NoStop}%
\bibitem [{\citenamefont {Chen}\ \emph {et~al.}(2018)\citenamefont {Chen},
  \citenamefont {Lee}, \citenamefont {Lin}, \citenamefont {Chen},\ and\
  \citenamefont {Wu}}]{chen2018liquid}%
  \BibitemOpen
  \bibfield  {author} {\bibinfo {author} {\bibfnamefont {H.-W.}\ \bibnamefont
  {Chen}}, \bibinfo {author} {\bibfnamefont {J.-H.}\ \bibnamefont {Lee}},
  \bibinfo {author} {\bibfnamefont {B.-Y.}\ \bibnamefont {Lin}}, \bibinfo
  {author} {\bibfnamefont {S.}~\bibnamefont {Chen}}, \ and\ \bibinfo {author}
  {\bibfnamefont {S.-T.}\ \bibnamefont {Wu}},\ }\href@noop {} {\bibfield
  {journal} {\bibinfo  {journal} {Light Sci. Appl.}\ }\textbf {\bibinfo
  {volume} {7}},\ \bibinfo {pages} {17168} (\bibinfo {year}
  {2018})}\BibitemShut {NoStop}%
\bibitem [{\citenamefont {Huang}\ \emph {et~al.}(2018)\citenamefont {Huang},
  \citenamefont {Liao}, \citenamefont {Chen},\ and\ \citenamefont
  {Wu}}]{huang2018liquid}%
  \BibitemOpen
  \bibfield  {author} {\bibinfo {author} {\bibfnamefont {Y.}~\bibnamefont
  {Huang}}, \bibinfo {author} {\bibfnamefont {E.}~\bibnamefont {Liao}},
  \bibinfo {author} {\bibfnamefont {R.}~\bibnamefont {Chen}}, \ and\ \bibinfo
  {author} {\bibfnamefont {S.-T.}\ \bibnamefont {Wu}},\ }\href@noop {}
  {\bibfield  {journal} {\bibinfo  {journal} {Appl. Sci.}\ }\textbf {\bibinfo
  {volume} {8}},\ \bibinfo {pages} {2366} (\bibinfo {year} {2018})}\BibitemShut
  {NoStop}%
\bibitem [{\citenamefont {Dabrowski}\ \emph {et~al.}(2018)\citenamefont
  {Dabrowski}, \citenamefont {Dziaduszek}, \citenamefont {Bozetka},
  \citenamefont {Piecek}, \citenamefont {Mazur}, \citenamefont {Chrunik},
  \citenamefont {Perkowski}, \citenamefont {Mrukiewicz}, \citenamefont
  {{\.Z}urowska},\ and\ \citenamefont {Weglowska}}]{dabrowski2018fluorinated}%
  \BibitemOpen
  \bibfield  {author} {\bibinfo {author} {\bibfnamefont {R.}~\bibnamefont
  {Dabrowski}}, \bibinfo {author} {\bibfnamefont {J.}~\bibnamefont
  {Dziaduszek}}, \bibinfo {author} {\bibfnamefont {J.}~\bibnamefont {Bozetka}},
  \bibinfo {author} {\bibfnamefont {W.}~\bibnamefont {Piecek}}, \bibinfo
  {author} {\bibfnamefont {R.}~\bibnamefont {Mazur}}, \bibinfo {author}
  {\bibfnamefont {M.}~\bibnamefont {Chrunik}}, \bibinfo {author} {\bibfnamefont
  {P.}~\bibnamefont {Perkowski}}, \bibinfo {author} {\bibfnamefont
  {M.}~\bibnamefont {Mrukiewicz}}, \bibinfo {author} {\bibfnamefont
  {M.}~\bibnamefont {{\.Z}urowska}}, \ and\ \bibinfo {author} {\bibfnamefont
  {D.}~\bibnamefont {Weglowska}},\ }\href@noop {} {\bibfield  {journal}
  {\bibinfo  {journal} {J. Mol. Liq.}\ }\textbf {\bibinfo {volume} {267}},\
  \bibinfo {pages} {415} (\bibinfo {year} {2018})}\BibitemShut {NoStop}%
\bibitem [{\citenamefont {Meyer}\ \emph {et~al.}(2020)\citenamefont {Meyer},
  \citenamefont {Blanc}, \citenamefont {Luckhurst}, \citenamefont {Davidson},\
  and\ \citenamefont {Dozov}}]{meyer2020biaxiality}%
  \BibitemOpen
  \bibfield  {author} {\bibinfo {author} {\bibfnamefont {C.}~\bibnamefont
  {Meyer}}, \bibinfo {author} {\bibfnamefont {C.}~\bibnamefont {Blanc}},
  \bibinfo {author} {\bibfnamefont {G.~R.}\ \bibnamefont {Luckhurst}}, \bibinfo
  {author} {\bibfnamefont {P.}~\bibnamefont {Davidson}}, \ and\ \bibinfo
  {author} {\bibfnamefont {I.}~\bibnamefont {Dozov}},\ }\href@noop {}
  {\bibfield  {journal} {\bibinfo  {journal} {Sci. Adv.}\ }\textbf {\bibinfo
  {volume} {6}},\ \bibinfo {pages} {eabb8212} (\bibinfo {year}
  {2020})}\BibitemShut {NoStop}%
\bibitem [{\citenamefont {Cousins}\ \emph {et~al.}(2019)\citenamefont
  {Cousins}, \citenamefont {Wilson}, \citenamefont {Mottram}, \citenamefont
  {Wilkes},\ and\ \citenamefont {Weegels}}]{cousins2019squeezing}%
  \BibitemOpen
  \bibfield  {author} {\bibinfo {author} {\bibfnamefont {J.}~\bibnamefont
  {Cousins}}, \bibinfo {author} {\bibfnamefont {S.}~\bibnamefont {Wilson}},
  \bibinfo {author} {\bibfnamefont {N.}~\bibnamefont {Mottram}}, \bibinfo
  {author} {\bibfnamefont {D.}~\bibnamefont {Wilkes}}, \ and\ \bibinfo {author}
  {\bibfnamefont {L.}~\bibnamefont {Weegels}},\ }\href@noop {} {\bibfield
  {journal} {\bibinfo  {journal} {Phys. Fluids}\ }\textbf {\bibinfo {volume}
  {31}},\ \bibinfo {pages} {083107} (\bibinfo {year} {2019})}\BibitemShut
  {NoStop}%
\bibitem [{\citenamefont {Luckhurst}\ and\ \citenamefont
  {Sluckin}(2015)}]{luckhurst2015biaxial}%
  \BibitemOpen
  \bibfield  {author} {\bibinfo {author} {\bibfnamefont {G.~R.}\ \bibnamefont
  {Luckhurst}}\ and\ \bibinfo {author} {\bibfnamefont {T.~J.}\ \bibnamefont
  {Sluckin}},\ }\href@noop {} {\emph {\bibinfo {title} {Biaxial nematic liquid
  crystals: theory, simulation and experiment}}}\ (\bibinfo  {publisher}
  {Wiley},\ \bibinfo {year} {2015})\BibitemShut {NoStop}%
\bibitem [{\citenamefont {Lee}\ \emph {et~al.}(2007)\citenamefont {Lee},
  \citenamefont {Lim}, \citenamefont {Kim},\ and\ \citenamefont
  {Jin}}]{lee2007dynamics}%
  \BibitemOpen
  \bibfield  {author} {\bibinfo {author} {\bibfnamefont {J.-H.}\ \bibnamefont
  {Lee}}, \bibinfo {author} {\bibfnamefont {T.-K.}\ \bibnamefont {Lim}},
  \bibinfo {author} {\bibfnamefont {W.-T.}\ \bibnamefont {Kim}}, \ and\
  \bibinfo {author} {\bibfnamefont {J.-I.}\ \bibnamefont {Jin}},\ }\href@noop
  {} {\bibfield  {journal} {\bibinfo  {journal} {J. Appl. Phys.}\ }\textbf
  {\bibinfo {volume} {101}},\ \bibinfo {pages} {034105} (\bibinfo {year}
  {2007})}\BibitemShut {NoStop}%
\bibitem [{\citenamefont {Berardi}\ and\ \citenamefont
  {Zannoni}(2008)}]{berardi2008}%
  \BibitemOpen
  \bibfield  {author} {\bibinfo {author} {\bibfnamefont {R.}~\bibnamefont
  {Berardi}}\ and\ \bibinfo {author} {\bibfnamefont {C.}~\bibnamefont
  {Zannoni}},\ }\href@noop {} {\bibfield  {journal} {\bibinfo  {journal} {J.
  Chem. Phys.}\ }\textbf {\bibinfo {volume} {113}},\ \bibinfo {pages} {5971}
  (\bibinfo {year} {2008})}\BibitemShut {NoStop}%
\bibitem [{\citenamefont {Ricci}\ \emph {et~al.}(2015)\citenamefont {Ricci},
  \citenamefont {Berardi},\ and\ \citenamefont {Zannoni}}]{ricci2015field}%
  \BibitemOpen
  \bibfield  {author} {\bibinfo {author} {\bibfnamefont {M.}~\bibnamefont
  {Ricci}}, \bibinfo {author} {\bibfnamefont {R.}~\bibnamefont {Berardi}}, \
  and\ \bibinfo {author} {\bibfnamefont {C.}~\bibnamefont {Zannoni}},\
  }\href@noop {} {\bibfield  {journal} {\bibinfo  {journal} {J. Chem. Phys.}\
  }\textbf {\bibinfo {volume} {143}},\ \bibinfo {pages} {084705} (\bibinfo
  {year} {2015})}\BibitemShut {NoStop}%
\bibitem [{\citenamefont {Taylor}\ and\ \citenamefont
  {Herzfeld}(1991)}]{taylor1991nematic}%
  \BibitemOpen
  \bibfield  {author} {\bibinfo {author} {\bibfnamefont {M.~P.}\ \bibnamefont
  {Taylor}}\ and\ \bibinfo {author} {\bibfnamefont {J.}~\bibnamefont
  {Herzfeld}},\ }\href@noop {} {\bibfield  {journal} {\bibinfo  {journal}
  {Phys. Rev. A.}\ }\textbf {\bibinfo {volume} {44}},\ \bibinfo {pages} {3742}
  (\bibinfo {year} {1991})}\BibitemShut {NoStop}%
\bibitem [{\citenamefont {Belli}\ \emph {et~al.}(2011)\citenamefont {Belli},
  \citenamefont {Patti}, \citenamefont {Dijkstra},\ and\ \citenamefont {van
  Roij}}]{belli2011polydispersity}%
  \BibitemOpen
  \bibfield  {author} {\bibinfo {author} {\bibfnamefont {S.}~\bibnamefont
  {Belli}}, \bibinfo {author} {\bibfnamefont {A.}~\bibnamefont {Patti}},
  \bibinfo {author} {\bibfnamefont {M.}~\bibnamefont {Dijkstra}}, \ and\
  \bibinfo {author} {\bibfnamefont {R.}~\bibnamefont {van Roij}},\ }\href@noop
  {} {\bibfield  {journal} {\bibinfo  {journal} {Phys. Rev. Lett.}\ }\textbf
  {\bibinfo {volume} {107}},\ \bibinfo {pages} {148303} (\bibinfo {year}
  {2011})}\BibitemShut {NoStop}%
\bibitem [{\citenamefont {Rafael}\ \emph {et~al.}(2020)\citenamefont {Rafael},
  \citenamefont {Cuetos}, \citenamefont {Corbett},\ and\ \citenamefont
  {Patti}}]{rafael2020self}%
  \BibitemOpen
  \bibfield  {author} {\bibinfo {author} {\bibfnamefont {E.~M.}\ \bibnamefont
  {Rafael}}, \bibinfo {author} {\bibfnamefont {A.}~\bibnamefont {Cuetos}},
  \bibinfo {author} {\bibfnamefont {D.}~\bibnamefont {Corbett}}, \ and\
  \bibinfo {author} {\bibfnamefont {A.}~\bibnamefont {Patti}},\ }\href@noop {}
  {\bibfield  {journal} {\bibinfo  {journal} {Soft Matter}\ }\textbf {\bibinfo
  {volume} {16}},\ \bibinfo {pages} {5565} (\bibinfo {year}
  {2020})}\BibitemShut {NoStop}%
\bibitem [{\citenamefont {Dussi}\ \emph {et~al.}(2018)\citenamefont {Dussi},
  \citenamefont {Tasios}, \citenamefont {Drwenski}, \citenamefont {van Roij},\
  and\ \citenamefont {Dijkstra}}]{dussi2018hard}%
  \BibitemOpen
  \bibfield  {author} {\bibinfo {author} {\bibfnamefont {S.}~\bibnamefont
  {Dussi}}, \bibinfo {author} {\bibfnamefont {N.}~\bibnamefont {Tasios}},
  \bibinfo {author} {\bibfnamefont {T.}~\bibnamefont {Drwenski}}, \bibinfo
  {author} {\bibfnamefont {R.}~\bibnamefont {van Roij}}, \ and\ \bibinfo
  {author} {\bibfnamefont {M.}~\bibnamefont {Dijkstra}},\ }\href@noop {}
  {\bibfield  {journal} {\bibinfo  {journal} {Phys. Rev. Lett.}\ }\textbf
  {\bibinfo {volume} {120}},\ \bibinfo {pages} {177801} (\bibinfo {year}
  {2018})}\BibitemShut {NoStop}%
\bibitem [{\citenamefont {Belli}\ \emph {et~al.}(2012)\citenamefont {Belli},
  \citenamefont {Dijkstra},\ and\ \citenamefont {van
  Roij}}]{belli2012depletion}%
  \BibitemOpen
  \bibfield  {author} {\bibinfo {author} {\bibfnamefont {S.}~\bibnamefont
  {Belli}}, \bibinfo {author} {\bibfnamefont {M.}~\bibnamefont {Dijkstra}}, \
  and\ \bibinfo {author} {\bibfnamefont {R.}~\bibnamefont {van Roij}},\
  }\href@noop {} {\bibfield  {journal} {\bibinfo  {journal} {J. Phys.: Condens.
  Matter}\ }\textbf {\bibinfo {volume} {24}},\ \bibinfo {pages} {284128}
  (\bibinfo {year} {2012})}\BibitemShut {NoStop}%
\bibitem [{\citenamefont {Leferink~op Reinink}\ \emph
  {et~al.}(2014)\citenamefont {Leferink~op Reinink}, \citenamefont {Belli},
  \citenamefont {van Roij}, \citenamefont {Dijkstra}, \citenamefont
  {Petukhov},\ and\ \citenamefont {Vroege}}]{op2014tuning}%
  \BibitemOpen
  \bibfield  {author} {\bibinfo {author} {\bibfnamefont {A.~B. G.~M.}\
  \bibnamefont {Leferink~op Reinink}}, \bibinfo {author} {\bibfnamefont
  {S.}~\bibnamefont {Belli}}, \bibinfo {author} {\bibfnamefont
  {R.}~\bibnamefont {van Roij}}, \bibinfo {author} {\bibfnamefont
  {M.}~\bibnamefont {Dijkstra}}, \bibinfo {author} {\bibfnamefont {A.~V.}\
  \bibnamefont {Petukhov}}, \ and\ \bibinfo {author} {\bibfnamefont {G.~J.}\
  \bibnamefont {Vroege}},\ }\href@noop {} {\bibfield  {journal} {\bibinfo
  {journal} {Soft Matter}\ }\textbf {\bibinfo {volume} {10}},\ \bibinfo {pages}
  {446} (\bibinfo {year} {2014})}\BibitemShut {NoStop}%
\bibitem [{\citenamefont {Cuetos}\ \emph {et~al.}(2019)\citenamefont {Cuetos},
  \citenamefont {Rafael}, \citenamefont {Corbett},\ and\ \citenamefont
  {Patti}}]{cuetos2019biaxial}%
  \BibitemOpen
  \bibfield  {author} {\bibinfo {author} {\bibfnamefont {A.}~\bibnamefont
  {Cuetos}}, \bibinfo {author} {\bibfnamefont {E.~M.}\ \bibnamefont {Rafael}},
  \bibinfo {author} {\bibfnamefont {D.}~\bibnamefont {Corbett}}, \ and\
  \bibinfo {author} {\bibfnamefont {A.}~\bibnamefont {Patti}},\ }\href@noop {}
  {\bibfield  {journal} {\bibinfo  {journal} {Soft Matter}\ }\textbf {\bibinfo
  {volume} {15}},\ \bibinfo {pages} {1922} (\bibinfo {year}
  {2019})}\BibitemShut {NoStop}%
\bibitem [{\citenamefont {Cuetos}\ \emph {et~al.}(2017)\citenamefont {Cuetos},
  \citenamefont {Dennison}, \citenamefont {Masters},\ and\ \citenamefont
  {Patti}}]{cuetos2017phase}%
  \BibitemOpen
  \bibfield  {author} {\bibinfo {author} {\bibfnamefont {A.}~\bibnamefont
  {Cuetos}}, \bibinfo {author} {\bibfnamefont {M.}~\bibnamefont {Dennison}},
  \bibinfo {author} {\bibfnamefont {A.}~\bibnamefont {Masters}}, \ and\
  \bibinfo {author} {\bibfnamefont {A.}~\bibnamefont {Patti}},\ }\href@noop {}
  {\bibfield  {journal} {\bibinfo  {journal} {Soft Matter}\ }\textbf {\bibinfo
  {volume} {13}},\ \bibinfo {pages} {4720} (\bibinfo {year}
  {2017})}\BibitemShut {NoStop}%
\bibitem [{\citenamefont {Patti}\ and\ \citenamefont
  {Cuetos}(2018)}]{patti2018monte}%
  \BibitemOpen
  \bibfield  {author} {\bibinfo {author} {\bibfnamefont {A.}~\bibnamefont
  {Patti}}\ and\ \bibinfo {author} {\bibfnamefont {A.}~\bibnamefont {Cuetos}},\
  }\href@noop {} {\bibfield  {journal} {\bibinfo  {journal} {Mol. Simul.}\
  }\textbf {\bibinfo {volume} {44}},\ \bibinfo {pages} {516} (\bibinfo {year}
  {2018})}\BibitemShut {NoStop}%
\bibitem [{\citenamefont {Yang}\ \emph {et~al.}(2018)\citenamefont {Yang},
  \citenamefont {Chen}, \citenamefont {Thanneeru}, \citenamefont {He},
  \citenamefont {Liu},\ and\ \citenamefont {Nie}}]{yang2018synthesis}%
  \BibitemOpen
  \bibfield  {author} {\bibinfo {author} {\bibfnamefont {Y.}~\bibnamefont
  {Yang}}, \bibinfo {author} {\bibfnamefont {G.}~\bibnamefont {Chen}}, \bibinfo
  {author} {\bibfnamefont {S.}~\bibnamefont {Thanneeru}}, \bibinfo {author}
  {\bibfnamefont {J.}~\bibnamefont {He}}, \bibinfo {author} {\bibfnamefont
  {K.}~\bibnamefont {Liu}}, \ and\ \bibinfo {author} {\bibfnamefont
  {Z.}~\bibnamefont {Nie}},\ }\href@noop {} {\bibfield  {journal} {\bibinfo
  {journal} {Nat. Commun.}\ }\textbf {\bibinfo {volume} {9}},\ \bibinfo {pages}
  {4513} (\bibinfo {year} {2018})}\BibitemShut {NoStop}%
\bibitem [{\citenamefont {Chiappini}\ \emph {et~al.}(2019)\citenamefont
  {Chiappini}, \citenamefont {Drwenski}, \citenamefont {Van~Roij},\ and\
  \citenamefont {Dijkstra}}]{chiappini2019biaxial}%
  \BibitemOpen
  \bibfield  {author} {\bibinfo {author} {\bibfnamefont {M.}~\bibnamefont
  {Chiappini}}, \bibinfo {author} {\bibfnamefont {T.}~\bibnamefont {Drwenski}},
  \bibinfo {author} {\bibfnamefont {R.}~\bibnamefont {Van~Roij}}, \ and\
  \bibinfo {author} {\bibfnamefont {M.}~\bibnamefont {Dijkstra}},\ }\href@noop
  {} {\bibfield  {journal} {\bibinfo  {journal} {Phys. Rev. Lett.}\ }\textbf
  {\bibinfo {volume} {123}},\ \bibinfo {pages} {068001} (\bibinfo {year}
  {2019})}\BibitemShut {NoStop}%
\bibitem [{\citenamefont {Tasios}\ and\ \citenamefont
  {Dijkstra}(2017)}]{tasios2017simulation}%
  \BibitemOpen
  \bibfield  {author} {\bibinfo {author} {\bibfnamefont {N.}~\bibnamefont
  {Tasios}}\ and\ \bibinfo {author} {\bibfnamefont {M.}~\bibnamefont
  {Dijkstra}},\ }\href@noop {} {\bibfield  {journal} {\bibinfo  {journal} {J.
  Chem. Phys.}\ }\textbf {\bibinfo {volume} {146}},\ \bibinfo {pages} {144901}
  (\bibinfo {year} {2017})}\BibitemShut {NoStop}%
\bibitem [{\citenamefont {van~den Pol}\ \emph {et~al.}(2009)\citenamefont
  {van~den Pol}, \citenamefont {Petukhov}, \citenamefont {Thies-Weesie},
  \citenamefont {Byelov},\ and\ \citenamefont {Vroege}}]{van2009experimental}%
  \BibitemOpen
  \bibfield  {author} {\bibinfo {author} {\bibfnamefont {E.}~\bibnamefont
  {van~den Pol}}, \bibinfo {author} {\bibfnamefont {A.}~\bibnamefont
  {Petukhov}}, \bibinfo {author} {\bibfnamefont {D.}~\bibnamefont
  {Thies-Weesie}}, \bibinfo {author} {\bibfnamefont {D.}~\bibnamefont
  {Byelov}}, \ and\ \bibinfo {author} {\bibfnamefont {G.}~\bibnamefont
  {Vroege}},\ }\href@noop {} {\bibfield  {journal} {\bibinfo  {journal} {Phys.
  Rev. Lett.}\ }\textbf {\bibinfo {volume} {103}},\ \bibinfo {pages} {258301}
  (\bibinfo {year} {2009})}\BibitemShut {NoStop}%
\bibitem [{\citenamefont {Peroukidis}\ \emph {et~al.}(2013)\citenamefont
  {Peroukidis}, \citenamefont {Vanakaras},\ and\ \citenamefont
  {Photinos}}]{peroukidis2013supramolecular}%
  \BibitemOpen
  \bibfield  {author} {\bibinfo {author} {\bibfnamefont {S.~D.}\ \bibnamefont
  {Peroukidis}}, \bibinfo {author} {\bibfnamefont {A.~G.}\ \bibnamefont
  {Vanakaras}}, \ and\ \bibinfo {author} {\bibfnamefont {D.~J.}\ \bibnamefont
  {Photinos}},\ }\href@noop {} {\bibfield  {journal} {\bibinfo  {journal}
  {Phys. Rev. E}\ }\textbf {\bibinfo {volume} {88}},\ \bibinfo {pages} {062508}
  (\bibinfo {year} {2013})}\BibitemShut {NoStop}%
\bibitem [{\citenamefont {Peroukidis}\ and\ \citenamefont
  {Vanakaras}(2013)}]{peroukidis2013phase}%
  \BibitemOpen
  \bibfield  {author} {\bibinfo {author} {\bibfnamefont {S.~D.}\ \bibnamefont
  {Peroukidis}}\ and\ \bibinfo {author} {\bibfnamefont {A.~G.}\ \bibnamefont
  {Vanakaras}},\ }\href@noop {} {\bibfield  {journal} {\bibinfo  {journal}
  {Soft Matter}\ }\textbf {\bibinfo {volume} {9}},\ \bibinfo {pages} {7419}
  (\bibinfo {year} {2013})}\BibitemShut {NoStop}%
\bibitem [{\citenamefont {Peroukidis}(2014)}]{peroukidis2014biaxial}%
  \BibitemOpen
  \bibfield  {author} {\bibinfo {author} {\bibfnamefont {S.~D.}\ \bibnamefont
  {Peroukidis}},\ }\href@noop {} {\bibfield  {journal} {\bibinfo  {journal}
  {Soft Matter}\ }\textbf {\bibinfo {volume} {10}},\ \bibinfo {pages} {4199}
  (\bibinfo {year} {2014})}\BibitemShut {NoStop}%
\bibitem [{\citenamefont {Skutnik}\ \emph {et~al.}(2020)\citenamefont
  {Skutnik}, \citenamefont {Geier},\ and\ \citenamefont
  {Schoen}}]{skutnik2020biaxial}%
  \BibitemOpen
  \bibfield  {author} {\bibinfo {author} {\bibfnamefont {R.~A.}\ \bibnamefont
  {Skutnik}}, \bibinfo {author} {\bibfnamefont {I.~S.}\ \bibnamefont {Geier}},
  \ and\ \bibinfo {author} {\bibfnamefont {M.}~\bibnamefont {Schoen}},\
  }\href@noop {} {\bibfield  {journal} {\bibinfo  {journal} {Mol. Phys.}\ ,\
  \bibinfo {pages} {1}} (\bibinfo {year} {2020})}\BibitemShut {NoStop}%
\bibitem [{\citenamefont {Van~den Pol}\ \emph
  {et~al.}(2010{\natexlab{a}})\citenamefont {Van~den Pol}, \citenamefont
  {Lupascu}, \citenamefont {Davidson},\ and\ \citenamefont
  {Vroege}}]{van2010isotropic}%
  \BibitemOpen
  \bibfield  {author} {\bibinfo {author} {\bibfnamefont {E.}~\bibnamefont
  {Van~den Pol}}, \bibinfo {author} {\bibfnamefont {A.}~\bibnamefont
  {Lupascu}}, \bibinfo {author} {\bibfnamefont {P.}~\bibnamefont {Davidson}}, \
  and\ \bibinfo {author} {\bibfnamefont {G.}~\bibnamefont {Vroege}},\
  }\href@noop {} {\bibfield  {journal} {\bibinfo  {journal} {J. Chem. Phys.}\
  }\textbf {\bibinfo {volume} {133}},\ \bibinfo {pages} {164504} (\bibinfo
  {year} {2010}{\natexlab{a}})}\BibitemShut {NoStop}%
\bibitem [{\citenamefont {Van~den Pol}\ \emph
  {et~al.}(2010{\natexlab{b}})\citenamefont {Van~den Pol}, \citenamefont
  {Lupascu}, \citenamefont {Diaconeasa}, \citenamefont {Petukhov},
  \citenamefont {Byelov},\ and\ \citenamefont {Vroege}}]{van2010onsager}%
  \BibitemOpen
  \bibfield  {author} {\bibinfo {author} {\bibfnamefont {E.}~\bibnamefont
  {Van~den Pol}}, \bibinfo {author} {\bibfnamefont {A.}~\bibnamefont
  {Lupascu}}, \bibinfo {author} {\bibfnamefont {M.}~\bibnamefont {Diaconeasa}},
  \bibinfo {author} {\bibfnamefont {A.}~\bibnamefont {Petukhov}}, \bibinfo
  {author} {\bibfnamefont {D.}~\bibnamefont {Byelov}}, \ and\ \bibinfo {author}
  {\bibfnamefont {G.}~\bibnamefont {Vroege}},\ }\href@noop {} {\bibfield
  {journal} {\bibinfo  {journal} {Chem. Phys. Lett.}\ }\textbf {\bibinfo
  {volume} {1}},\ \bibinfo {pages} {2174} (\bibinfo {year}
  {2010}{\natexlab{b}})}\BibitemShut {NoStop}%
\bibitem [{\citenamefont {Peroukidis}\ \emph {et~al.}(2020)\citenamefont
  {Peroukidis}, \citenamefont {Klapp},\ and\ \citenamefont
  {Vanakaras}}]{peroukidis2020field}%
  \BibitemOpen
  \bibfield  {author} {\bibinfo {author} {\bibfnamefont {S.~D.}\ \bibnamefont
  {Peroukidis}}, \bibinfo {author} {\bibfnamefont {S.~H.}\ \bibnamefont
  {Klapp}}, \ and\ \bibinfo {author} {\bibfnamefont {A.~G.}\ \bibnamefont
  {Vanakaras}},\ }\href@noop {} {\bibfield  {journal} {\bibinfo  {journal}
  {Soft Matter}\ }\textbf {\bibinfo {volume} {16}},\ \bibinfo {pages} {10667}
  (\bibinfo {year} {2020})}\BibitemShut {NoStop}%
\bibitem [{\citenamefont {Cuetos}\ \emph {et~al.}(2008)\citenamefont {Cuetos},
  \citenamefont {Galindo},\ and\ \citenamefont
  {Jackson}}]{cuetos2008thermotropic}%
  \BibitemOpen
  \bibfield  {author} {\bibinfo {author} {\bibfnamefont {A.}~\bibnamefont
  {Cuetos}}, \bibinfo {author} {\bibfnamefont {A.}~\bibnamefont {Galindo}}, \
  and\ \bibinfo {author} {\bibfnamefont {G.}~\bibnamefont {Jackson}},\
  }\href@noop {} {\bibfield  {journal} {\bibinfo  {journal} {Phys. Rev. Lett.}\
  }\textbf {\bibinfo {volume} {101}},\ \bibinfo {pages} {237802} (\bibinfo
  {year} {2008})}\BibitemShut {NoStop}%
\bibitem [{\citenamefont {Patti}\ and\ \citenamefont
  {Cuetos}(2012)}]{patti2012brownian}%
  \BibitemOpen
  \bibfield  {author} {\bibinfo {author} {\bibfnamefont {A.}~\bibnamefont
  {Patti}}\ and\ \bibinfo {author} {\bibfnamefont {A.}~\bibnamefont {Cuetos}},\
  }\href@noop {} {\bibfield  {journal} {\bibinfo  {journal} {Phys. Rev. E}\
  }\textbf {\bibinfo {volume} {86}},\ \bibinfo {pages} {011403} (\bibinfo
  {year} {2012})}\BibitemShut {NoStop}%
\bibitem [{\citenamefont {Cuetos}\ and\ \citenamefont
  {Patti}(2015)}]{cuetos2015equivalence}%
  \BibitemOpen
  \bibfield  {author} {\bibinfo {author} {\bibfnamefont {A.}~\bibnamefont
  {Cuetos}}\ and\ \bibinfo {author} {\bibfnamefont {A.}~\bibnamefont {Patti}},\
  }\href@noop {} {\bibfield  {journal} {\bibinfo  {journal} {Phys. Rev. E}\
  }\textbf {\bibinfo {volume} {92}},\ \bibinfo {pages} {022302} (\bibinfo
  {year} {2015})}\BibitemShut {NoStop}%
\bibitem [{\citenamefont {Corbett}\ \emph {et~al.}(2018)\citenamefont
  {Corbett}, \citenamefont {Cuetos}, \citenamefont {Dennison},\ and\
  \citenamefont {Patti}}]{corbett2018dynamic}%
  \BibitemOpen
  \bibfield  {author} {\bibinfo {author} {\bibfnamefont {D.}~\bibnamefont
  {Corbett}}, \bibinfo {author} {\bibfnamefont {A.}~\bibnamefont {Cuetos}},
  \bibinfo {author} {\bibfnamefont {M.}~\bibnamefont {Dennison}}, \ and\
  \bibinfo {author} {\bibfnamefont {A.}~\bibnamefont {Patti}},\ }\href@noop {}
  {\bibfield  {journal} {\bibinfo  {journal} {Phys. Chem. Chem. Phys.}\
  }\textbf {\bibinfo {volume} {20}},\ \bibinfo {pages} {15118} (\bibinfo {year}
  {2018})}\BibitemShut {NoStop}%
\bibitem [{\citenamefont {Daza}\ \emph {et~al.}(2020)\citenamefont {Daza},
  \citenamefont {Cuetos},\ and\ \citenamefont {Patti}}]{daza2020dynamic}%
  \BibitemOpen
  \bibfield  {author} {\bibinfo {author} {\bibfnamefont {F.~A.~G.}\
  \bibnamefont {Daza}}, \bibinfo {author} {\bibfnamefont {A.}~\bibnamefont
  {Cuetos}}, \ and\ \bibinfo {author} {\bibfnamefont {A.}~\bibnamefont
  {Patti}},\ }\href@noop {} {\bibfield  {journal} {\bibinfo  {journal} {Phys.
  Rev. E}\ }\textbf {\bibinfo {volume} {102}},\ \bibinfo {pages} {013302}
  (\bibinfo {year} {2020})}\BibitemShut {NoStop}%
\bibitem [{\citenamefont {Chiappini}\ \emph
  {et~al.}(2020{\natexlab{a}})\citenamefont {Chiappini}, \citenamefont
  {Patti},\ and\ \citenamefont {Dijkstra}}]{chiappini2020}%
  \BibitemOpen
  \bibfield  {author} {\bibinfo {author} {\bibfnamefont {M.}~\bibnamefont
  {Chiappini}}, \bibinfo {author} {\bibfnamefont {A.}~\bibnamefont {Patti}}, \
  and\ \bibinfo {author} {\bibfnamefont {M.}~\bibnamefont {Dijkstra}},\ }\href
  {\doibase 10.1103/PhysRevE.102.040601} {\bibfield  {journal} {\bibinfo
  {journal} {Phys. Rev. E}\ }\textbf {\bibinfo {volume} {102}},\ \bibinfo
  {pages} {040601} (\bibinfo {year} {2020}{\natexlab{a}})}\BibitemShut
  {NoStop}%
\bibitem [{\citenamefont {Cuetos}\ and\ \citenamefont
  {Patti}(2020)}]{cuetos2020dynamics}%
  \BibitemOpen
  \bibfield  {author} {\bibinfo {author} {\bibfnamefont {A.}~\bibnamefont
  {Cuetos}}\ and\ \bibinfo {author} {\bibfnamefont {A.}~\bibnamefont {Patti}},\
  }\href@noop {} {\bibfield  {journal} {\bibinfo  {journal} {Phys. Rev. E}\
  }\textbf {\bibinfo {volume} {101}},\ \bibinfo {pages} {052702} (\bibinfo
  {year} {2020})}\BibitemShut {NoStop}%
\bibitem [{\citenamefont {Lebovka}\ \emph {et~al.}(2019)\citenamefont
  {Lebovka}, \citenamefont {Vygornitskii},\ and\ \citenamefont
  {Tarasevich}}]{levobka2019}%
  \BibitemOpen
  \bibfield  {author} {\bibinfo {author} {\bibfnamefont {N.~I.}\ \bibnamefont
  {Lebovka}}, \bibinfo {author} {\bibfnamefont {N.~V.}\ \bibnamefont
  {Vygornitskii}}, \ and\ \bibinfo {author} {\bibfnamefont {Y.~Y.}\
  \bibnamefont {Tarasevich}},\ }\href {\doibase 10.1103/PhysRevE.100.042139}
  {\bibfield  {journal} {\bibinfo  {journal} {Phys. Rev. E}\ }\textbf {\bibinfo
  {volume} {100}},\ \bibinfo {pages} {042139} (\bibinfo {year}
  {2019})}\BibitemShut {NoStop}%
\bibitem [{\citenamefont {Chiappini}\ \emph
  {et~al.}(2020{\natexlab{b}})\citenamefont {Chiappini}, \citenamefont
  {Grelet},\ and\ \citenamefont {Dijkstra}}]{chiappini_2020_2}%
  \BibitemOpen
  \bibfield  {author} {\bibinfo {author} {\bibfnamefont {M.}~\bibnamefont
  {Chiappini}}, \bibinfo {author} {\bibfnamefont {E.}~\bibnamefont {Grelet}}, \
  and\ \bibinfo {author} {\bibfnamefont {M.}~\bibnamefont {Dijkstra}},\ }\href
  {\doibase 10.1103/PhysRevLett.124.087801} {\bibfield  {journal} {\bibinfo
  {journal} {Phys. Rev. Lett.}\ }\textbf {\bibinfo {volume} {124}},\ \bibinfo
  {pages} {087801} (\bibinfo {year} {2020}{\natexlab{b}})}\BibitemShut
  {NoStop}%
\bibitem [{\citenamefont {Cuetos}\ \emph {et~al.}(2018)\citenamefont {Cuetos},
  \citenamefont {Morillo},\ and\ \citenamefont {Patti}}]{morillo2018}%
  \BibitemOpen
  \bibfield  {author} {\bibinfo {author} {\bibfnamefont {A.}~\bibnamefont
  {Cuetos}}, \bibinfo {author} {\bibfnamefont {N.}~\bibnamefont {Morillo}}, \
  and\ \bibinfo {author} {\bibfnamefont {A.}~\bibnamefont {Patti}},\ }\href
  {\doibase 10.1103/PhysRevE.98.042129} {\bibfield  {journal} {\bibinfo
  {journal} {Phys. Rev. E}\ }\textbf {\bibinfo {volume} {98}},\ \bibinfo
  {pages} {042129} (\bibinfo {year} {2018})}\BibitemShut {NoStop}%
\bibitem [{\citenamefont {Stannarius}(2008)}]{stannarius2008comment}%
  \BibitemOpen
  \bibfield  {author} {\bibinfo {author} {\bibfnamefont {R.}~\bibnamefont
  {Stannarius}},\ }\href@noop {} {\bibfield  {journal} {\bibinfo  {journal} {J.
  Appl. Phys.}\ }\textbf {\bibinfo {volume} {104}},\ \bibinfo {pages} {034105}
  (\bibinfo {year} {2008})}\BibitemShut {NoStop}%
\bibitem [{\citenamefont {Gottschalk}\ \emph {et~al.}(1996)\citenamefont
  {Gottschalk}, \citenamefont {Lin},\ and\ \citenamefont
  {Manocha}}]{gottschalk1996obbtree}%
  \BibitemOpen
  \bibfield  {author} {\bibinfo {author} {\bibfnamefont {S.}~\bibnamefont
  {Gottschalk}}, \bibinfo {author} {\bibfnamefont {M.~C.}\ \bibnamefont {Lin}},
  \ and\ \bibinfo {author} {\bibfnamefont {D.}~\bibnamefont {Manocha}},\
  }\href@noop {} {\bibfield  {journal} {\bibinfo  {journal} {Comp. Graph.}\
  }\textbf {\bibinfo {volume} {30}},\ \bibinfo {pages} {171} (\bibinfo {year}
  {1996})}\BibitemShut {NoStop}%
\bibitem [{\citenamefont {John}\ and\ \citenamefont
  {Escobedo}(2005)}]{john2005phase}%
  \BibitemOpen
  \bibfield  {author} {\bibinfo {author} {\bibfnamefont {B.~S.}\ \bibnamefont
  {John}}\ and\ \bibinfo {author} {\bibfnamefont {F.~A.}\ \bibnamefont
  {Escobedo}},\ }\href@noop {} {\bibfield  {journal} {\bibinfo  {journal} {J.
  Phys. Chem. B}\ }\textbf {\bibinfo {volume} {109}},\ \bibinfo {pages} {23008}
  (\bibinfo {year} {2005})}\BibitemShut {NoStop}%
\bibitem [{\citenamefont {John}\ \emph {et~al.}(2008)\citenamefont {John},
  \citenamefont {Juhlin},\ and\ \citenamefont {Escobedo}}]{john2008phase}%
  \BibitemOpen
  \bibfield  {author} {\bibinfo {author} {\bibfnamefont {B.~S.}\ \bibnamefont
  {John}}, \bibinfo {author} {\bibfnamefont {C.}~\bibnamefont {Juhlin}}, \ and\
  \bibinfo {author} {\bibfnamefont {F.~A.}\ \bibnamefont {Escobedo}},\
  }\href@noop {} {\bibfield  {journal} {\bibinfo  {journal} {J. Chem. Phys.}\
  }\textbf {\bibinfo {volume} {128}},\ \bibinfo {pages} {044909} (\bibinfo
  {year} {2008})}\BibitemShut {NoStop}%
\bibitem [{\citenamefont {Carrasco}\ and\ \citenamefont {de~la
  Torre}(1999)}]{carrasco1999hydrodynamic}%
  \BibitemOpen
  \bibfield  {author} {\bibinfo {author} {\bibfnamefont {B.}~\bibnamefont
  {Carrasco}}\ and\ \bibinfo {author} {\bibfnamefont {J.~G.}\ \bibnamefont
  {de~la Torre}},\ }\href@noop {} {\bibfield  {journal} {\bibinfo  {journal}
  {Biophys. J.}\ }\textbf {\bibinfo {volume} {76}},\ \bibinfo {pages} {3044}
  (\bibinfo {year} {1999})}\BibitemShut {NoStop}%
\bibitem [{\citenamefont {Garc{\'\i}a de~la Torre}\ \emph
  {et~al.}(2007)\citenamefont {Garc{\'\i}a de~la Torre}, \citenamefont {del
  Rio~Echenique},\ and\ \citenamefont {Ortega}}]{garcia2007improved}%
  \BibitemOpen
  \bibfield  {author} {\bibinfo {author} {\bibfnamefont {J.}~\bibnamefont
  {Garc{\'\i}a de~la Torre}}, \bibinfo {author} {\bibfnamefont
  {G.}~\bibnamefont {del Rio~Echenique}}, \ and\ \bibinfo {author}
  {\bibfnamefont {A.}~\bibnamefont {Ortega}},\ }\href@noop {} {\bibfield
  {journal} {\bibinfo  {journal} {J. Phys. Chem. B}\ }\textbf {\bibinfo
  {volume} {111}},\ \bibinfo {pages} {955} (\bibinfo {year}
  {2007})}\BibitemShut {NoStop}%
\bibitem [{\citenamefont {Iglewicz}\ and\ \citenamefont
  {Hoaglin}(1993)}]{iglewicz1993detect}%
  \BibitemOpen
  \bibfield  {author} {\bibinfo {author} {\bibfnamefont {B.}~\bibnamefont
  {Iglewicz}}\ and\ \bibinfo {author} {\bibfnamefont {D.~C.}\ \bibnamefont
  {Hoaglin}},\ }\href@noop {} {\emph {\bibinfo {title} {How to detect and
  handle outliers}}},\ Vol.~\bibinfo {volume} {16}\ (\bibinfo  {publisher} {Asq
  Press},\ \bibinfo {year} {1993})\BibitemShut {NoStop}%
\bibitem [{\citenamefont {Mazza}\ \emph {et~al.}(2006)\citenamefont {Mazza},
  \citenamefont {Giovambattista}, \citenamefont {Starr},\ and\ \citenamefont
  {Stanley}}]{mazza2006relation}%
  \BibitemOpen
  \bibfield  {author} {\bibinfo {author} {\bibfnamefont {M.~G.}\ \bibnamefont
  {Mazza}}, \bibinfo {author} {\bibfnamefont {N.}~\bibnamefont
  {Giovambattista}}, \bibinfo {author} {\bibfnamefont {F.~W.}\ \bibnamefont
  {Starr}}, \ and\ \bibinfo {author} {\bibfnamefont {H.~E.}\ \bibnamefont
  {Stanley}},\ }\href@noop {} {\bibfield  {journal} {\bibinfo  {journal} {Phys.
  Rev. Lett.}\ }\textbf {\bibinfo {volume} {96}},\ \bibinfo {pages} {057803}
  (\bibinfo {year} {2006})}\BibitemShut {NoStop}%
\bibitem [{\citenamefont {Mazza}\ \emph {et~al.}(2007)\citenamefont {Mazza},
  \citenamefont {Giovambattista}, \citenamefont {Stanley},\ and\ \citenamefont
  {Starr}}]{mazza2007connection}%
  \BibitemOpen
  \bibfield  {author} {\bibinfo {author} {\bibfnamefont {M.~G.}\ \bibnamefont
  {Mazza}}, \bibinfo {author} {\bibfnamefont {N.}~\bibnamefont
  {Giovambattista}}, \bibinfo {author} {\bibfnamefont {H.~E.}\ \bibnamefont
  {Stanley}}, \ and\ \bibinfo {author} {\bibfnamefont {F.~W.}\ \bibnamefont
  {Starr}},\ }\href@noop {} {\bibfield  {journal} {\bibinfo  {journal} {Phys.
  Rev. E}\ }\textbf {\bibinfo {volume} {76}},\ \bibinfo {pages} {031203}
  (\bibinfo {year} {2007})}\BibitemShut {NoStop}%
\end{thebibliography}%

\end{document}